\documentclass[amsmath,twocolumn,amssymb,superscriptaddress,floatfix]{revtex4-1}

\usepackage{graphicx}
\usepackage{amssymb}
\usepackage{epstopdf}
\relax
\begin{document}

\title{Identification of a possible  superconducting transition above
room temperature in natural graphite crystals}

\author{Christian E. Precker}
\address{Division of Superconductivity and Magnetism, Institut
f\"ur Experimentelle Physik II, Universit\"{a}t Leipzig,
Linn\'{e}stra{\ss}e 5, D-04103 Leipzig, Germany} \author{Pablo D.
Esquinazi} \address{Division of Superconductivity and Magnetism,
Institut f\"ur Experimentelle Physik II, Universit\"{a}t Leipzig,
Linn\'{e}stra{\ss}e 5, D-04103 Leipzig, Germany}
\author{Ana Champi}
 \address{Centro de Ci\^{e}ncias Naturais e Humanas, Universidade Federal do
ABC, 09210-170, Santo Andr\'{e}, S{\~a}o Paulo, Brazil}
\author{Jos\'e Barzola-Quiquia}
\address{Division of Superconductivity and Magnetism, Institut
f\"ur Experimentelle Physik II, Universit\"{a}t Leipzig,
Linn\'{e}stra{\ss}e 5, D-04103 Leipzig, Germany}
\author{Mahsa Zoraghi}
\address{Division of Superconductivity and Magnetism, Institut
f\"ur Experimentelle Physik II, Universit\"{a}t Leipzig,
Linn\'{e}stra{\ss}e 5, D-04103 Leipzig, Germany}
\author{Santiago Mui{\~n}os-Landin}
\address{Division of Superconductivity and Magnetism, Institut
f\"ur Experimentelle Physik II, Universit\"{a}t Leipzig,
Linn\'{e}stra{\ss}e 5, D-04103 Leipzig, Germany}
\author{Annette Setzer}
\address{Division of Superconductivity and Magnetism, Institut
f\"ur Experimentelle Physik II, Universit\"{a}t Leipzig,
Linn\'{e}stra{\ss}e 5, D-04103 Leipzig, Germany}
\author{Winfried B\"ohlmann}
\address{Division of Superconductivity and Magnetism, Institut
f\"ur Experimentelle Physik II, Universit\"{a}t Leipzig,
Linn\'{e}stra{\ss}e 5, D-04103 Leipzig, Germany}
\author{Daniel Spemann}\altaffiliation[Present address: ]{Leibniz Institute of Surface
Modification, Physical Department, Permoserstr. 15, 04318 Leipzig,
Germany} \address{Division of Nuclear Solid State Physics,
Institut f\"ur Experimentelle Physik II, Universit\"{a}t Leipzig,
Linn\'{e}stra{\ss}e 5, D-04103 Leipzig, Germany}
\author{Jan Meijer}
\address{Division of Nuclear Solid State Physics, Institut
f\"ur Experimentelle Physik II, Universit\"{a}t Leipzig,
Linn\'{e}stra{\ss}e 5, D-04103 Leipzig, Germany}
\author{Tom Muenster}
\address{Institut f\"ur Mineralogie, Kristallographie und
Materialwissenschaft, Fakult\"at f\"ur Chemie und Mineralogie,
Universit\"{a}t Leipzig, Scharnhorststra{\ss}e 20, D-04275
Leipzig, Germany}
\author{Oliver Baehre}
\address{Institut f\"ur Mineralogie, Kristallographie und
Materialwissenschaft, Fakult\"at f\"ur Chemie und Mineralogie,
Universit\"{a}t Leipzig, Scharnhorststra{\ss}e 20, D-04275
Leipzig, Germany}
\author{Gert Kloess}
\address{Institut f\"ur Mineralogie, Kristallographie und
Materialwissenschaft, Fakult\"at f\"ur Chemie und Mineralogie,
Universit\"{a}t Leipzig, Scharnhorststra{\ss}e 20, D-04275
Leipzig, Germany}
\author{Henning Beth}
\address{Golden Bowerbird Pty Ltd., 32 Alidenes Rd., 2482
Mullumbimby, NSW Australia}


\begin{abstract}
Measuring  with high precision the electrical resistance of
 highly ordered natural graphite samples from a Brazil mine, we
have identified a transition at $\sim$350~K with $\sim$40~K
transition width. The step-like change in temperature of the
resistance, its magnetic irreversibility and time dependence after
a field change, consistent with trapped flux and flux creep, and
the partial magnetic flux expulsion obtained by magnetization
measurements, suggest the existence of granular superconductivity
below 350~K. The zero-field virgin state can only be reached again
after zero field cooling the sample from above the transition.
Paradoxically, the extraordinarily high transition temperature we
found for this and several other graphite samples is the reason
why this transition remained undetected so far. The existence of
well ordered rhombohedral graphite phase in all measured samples
has been proved by x-rays diffraction measurements, suggesting its
interfaces with the Bernal phase as a possible origin for the
high-temperature superconductivity, as theoretical studies
predicted. The localization of the granular superconductivity at
these two dimensional  interfaces prevents  the observation of a
zero resistance state or of a full Meissner state.
\end{abstract}

\maketitle

\section{Introduction}

The recently reported record temperature for superconductivity at
203~K in the sulfur hydride system at high pressures\cite{dro15}
appears to be consistent with the Bardeen--Cooper--Schrieffer
(BCS) theory for conventional superconductivity. Theoretical
estimates suggest the proximity of the Fermi surface to a Van Hove
singularity as one possible reason for the high-temperature
superconductivity\cite{bia15}. In this case the interaction
between carriers\cite{yud14} would give rise to the
Khodel-Shaginyan flat band, a band with a dispersionless energy
relation\cite{kho90-2,vol94}. Theoretical studies predict
superconductivity at high temperatures in electronic systems with
a flat band\cite{vol13,peo15} as, for example, at the surface of
rhombohedral graphite\cite{kop13} or at the interfaces between
this and Bernal graphite\cite{mun13}. Whereas evidence for flat
bands at the surface of epitaxial rhombohedral multilayer graphene
was recently found\cite{pie15}, hints for the existence of
superconductivity above room temperature in different
graphite-based systems were reported in the last 40
years\cite{ant74,ant75,yakovjltp00,esqpip,kaw13} without
providing, however, any knowledge of the critical temperature
$T_c$.

The recently observed superconductivity at $T_c \sim 5~$K in
intercalated Ca and Li-doped graphene\cite{cha15,lud15,ich16} is
probably related to the increase of the density of states due to a
change in the carrier density. The low critical temperature is due
to the common exponential suppression from the BCS equation for
systems with a quadratic dispersion relation. In case of having
similar Cooper pairs interaction strength and within the flat-band
predictions, one expects  a critical temperature $T_c$ orders of
magnitude larger because in this case  $T_c$ is {\em proportional}
to the pairing interaction strength and to the area of the flat
band in momentum space. Therefore, for a finite electron pairing
interaction strength, we  expect that at certain interfaces
between rhombohedral and Bernal stacking ordered
regions\cite{kop13,mun13}, or at certain paths between
(small-angle) twisted Bernal crystalline
regions\cite{San-Jose2013,esqarx14}, superconductivity may be
triggered in graphite with a considerably higher critical
temperature than in most bulk superconductors. The transition in
the electrical resistance we found at $T \gtrsim 350~$K in several
graphite samples and the behavior of the trapped flux and partial
flux expulsion vs. temperature suggest that at that temperature
superconducting regions are formed. Lowering temperature the size
of these regions may increase as well as their Josephson coupling
triggering the large decrease in the resistance below $\sim 100~$K
observed in graphite samples with interfaces. In this paper we
present  results obtained from different graphite samples, natural
graphite crystals from Brazil and Sri Lanka, thin flakes obtained
from those bulk crystals, and bulk and thin flakes from commercial
highly oriented pyrolytic graphite (HOPG) samples.

\section{Experimental details}
\label{details}

The images of the internal structure of the natural graphite
samples were obtained  using the scanning transmission electron
microscope (STEM)  of a Nova NanoLab 200 dual beam microscope from
the FEI company (Eindhoven). The lamellae were prepared with STEM
using the in-situ lift out method of the microscope cutting them
normal to the graphene layers. The STEM pictures were done with a
voltage applied to the electron column of 20~kV and the currents
used were between 38 to 140~pA.

The impurity content of the samples was determined by RBS/PIXE
using a 2~MeV proton beam of 0.8~mm diameter. From the surface of
the samples we removed the first hundreds of nanometers  that
usually have some contamination. The PIXE measurements of the
clear surface and with a penetration depth of $\sim 35~\mu$m
indicate a concentration of $6.4~$ppm of Fe and 5.9~ppm of Ti;
most of the other elements being below the detection limit. Here
``ppm" means $\mu$g element per gram sample. A thorough study of
the impurity concentration of the HOPG sample used in this study
(Advanced Ceramics, grade A) has been recently
published\cite{spe14}. The total magnetic elements concentration
in the HOPG sample used in this work is below 2~ppm.

The electrical resistance measurements were done in a He-flow
cryostat from Quantum Design with a superconducting solenoid. The
field provided by the solenoid in the low field region as well as
the rest field, at nominal zero field, was measured with a Hall
sensor especially designed to measure low fields. After applying
fields below 100~mT the rest field trapped by the solenoid at the
sample position and at nominally zero field, was below 0.1~mT.

We used the usual four-contacts method for the resistance
measurements, two for the current across the whole sample width
and flowing parallel to the graphene layers and interfaces, and
two for the voltage. The voltage electrodes were placed always on
the top graphene layer of the samples. To check for the
reproducibility of the observed behavior, the voltage electrodes
were placed contacting half of the sample width as well as the
whole width. The high-resolution resistance measurements were done
using an AC bridge (16~Hz) LR-700 with
 $10^{-5}$ relative resolution. Due to the low resistance (m$\Omega$) value
of some of the bulk samples, the input currents ranged from 0.1 to
30~mA. The shown results did not depend on the current value
within experimental error.

The temperature dependence of the resistance has been measured in
equilibrium avoiding any sweeping measuring mode, commonly used
for measurements in a broad temperature range. It means that after
reaching and regulating the selected temperature, we waited 5
minutes before the measurement started. This measuring mode is
time consuming but mandatory, especially to measure the tiny
transition at $T > 320~$K we describe below. The maximum
temperature available in the used equipment was 390~K.

The magnetization was measured using a superconducting quantum
interferometer device (SQUID) from Quantum Design. The used RSO
option allows us to measure the magnetic moments of our bulk
samples with a resolution $\lesssim 10^{-8}~$emu at low applied
fields.

\section{X-rays characterization of selected samples}

Ordered graphite samples can exhibit a regular pattern of
alternating Bernal (ABAB...) (2H) and rhombohedral (ABCABC...)
(3R) stacking with areas as small as (200~nm)$^2$ \cite{hat13}. A
certain amount of isolated 3R crystallites has been detected in
 bulk samples without suffering severe shear deformation \cite{lin12}.
 Upon graphite sample, a concentration of up to 30\% of the 3R
stacking order  is possible, especially in natural graphite
crystals\cite{kelly}. Therefore, in this work we studied the
transport properties and the internal structure of natural
graphite samples from Minas Gerais in Brazil. These highly ordered
natural graphite samples show in general an extraordinarily large
resistance ratio $R(300)/R(4) \gtrsim 20$, which is an indication
of highly conducting and large interfaces in agreement with
low-energy STEM results \cite{gar12}. For comparison, we also
measured thin flakes obtained from the same natural graphite
samples from Brazil, a bulk natural graphite sample from Sri Lanka
as well as bulk and thin flakes from HOPG of grade~A with 2H and
3R phases.

\begin{figure}
\begin{center}
\includegraphics[width=0.5\textwidth]{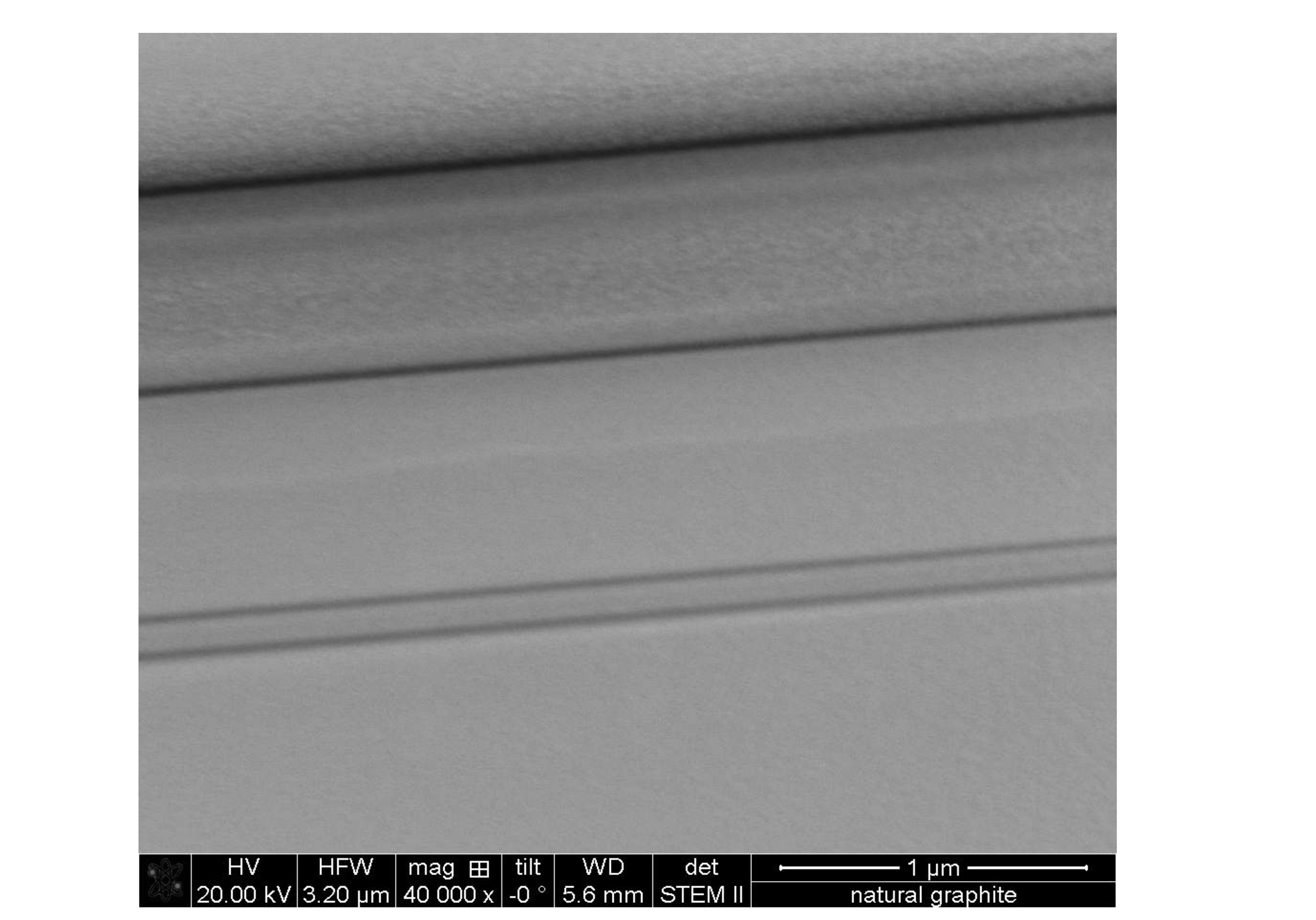}
\caption{ Internal structure of a 100~nm thick  lamella from a
Brazilian natural graphite  obtained with a transmission electron
microscope at low energy  and with the electron beam parallel to
the graphene planes. The bottom right bar indicates $1~\mu$m
length. Regions with different grey colors indicate either regions
with Bernal or rhombohedral stacking (with interfaces between
them) or twisted Bernal regions around the $c-$axis. The $c-$axis
is perpendicular to the graphene layers and interfaces.}
\label{TEM}
\end{center}
\end{figure}

The internal structure of the natural graphite flakes, revealed by
the STEM measurements, see Fig.~\ref{TEM}, consists of well
defined and several micrometers long interfaces between
crystalline regions with  different stacking orders or twisted
regions  around the $c-$axis. Clear evidence for the existence of
the 2H and 3R phases
 in the measured samples is
obtained
 by x-ray diffraction (XRD).

\begin{table}[]
\begin{tabular}{llll}
Hexagonal & & Rhombohedral & \\
\hline
$hkl$ &   2$\theta$(degree) &  $hkl$ &   2$\theta$(degree)\\
\hline
002      & 26.543 &  003& 26.611  \\
100      & 42.321 &  & \\
&& 101 & 43.454 \\
011      & 44.518 & & \\
&      & 012 & 46.321  \\
012      & 50.656 & & \\
004      & 54.662 & 006 & 54.812  \\
&      & 104 & 56.676  \\
013    & 59.852   && \\
&     & 015 &  63.677   \\
014     &71.471  && \\
110 & 77.400& 110 &77.696  \\
  & &107 & 80.738 \\
 112 & 83.527  &113 &83.848  \\
 015 &85.375  & &  \\
 006 &87.053 &009 &87.330 \\
 \hline
\end{tabular}
\caption{Expected Bragg angles for the hexagonal (Bernal, 2H) and
rhombohedral (3R) stacking orders in graphite for several Miller
indices ($hkl$). Note that there are several Bragg maxima not
suitable for distinguishing both stacking orders. This table is
taken from the Inorganic Crystal Structure Database (ICSD),
Fachinformationszentrum Karlsruhe (FIZ) \& National Institute of
Standards and Technology (NIST) (2005).} \label{table}
\end{table}

XRD studies were carried out in order to show the coexistence of
well ordered hexagonal and rhombohedral graphite in the
investigated samples. The XRD measurements of the natural graphite
samples and highly oriented pyrolytic graphite (HOPG) were
performed with a D8 Discover (Bruker AXS) operated at 40~mA and
40~kV (Cu-K$_\alpha$) and equipped with a Goebel mirror. The
measuring area on the samples was limited to 2~mm$^2$ using a
pinhole and a snout. The XRD measurements of  ultrapure graphite
powder RWA-T were done with a Phillips Xpert diffractometer.
Table~\ref{table} lists the expected Bragg angles for the Bernal
(2H) and rhombohedral (3R) stacking order phases.

\begin{figure}
\begin{center}
\includegraphics[width=1\linewidth]{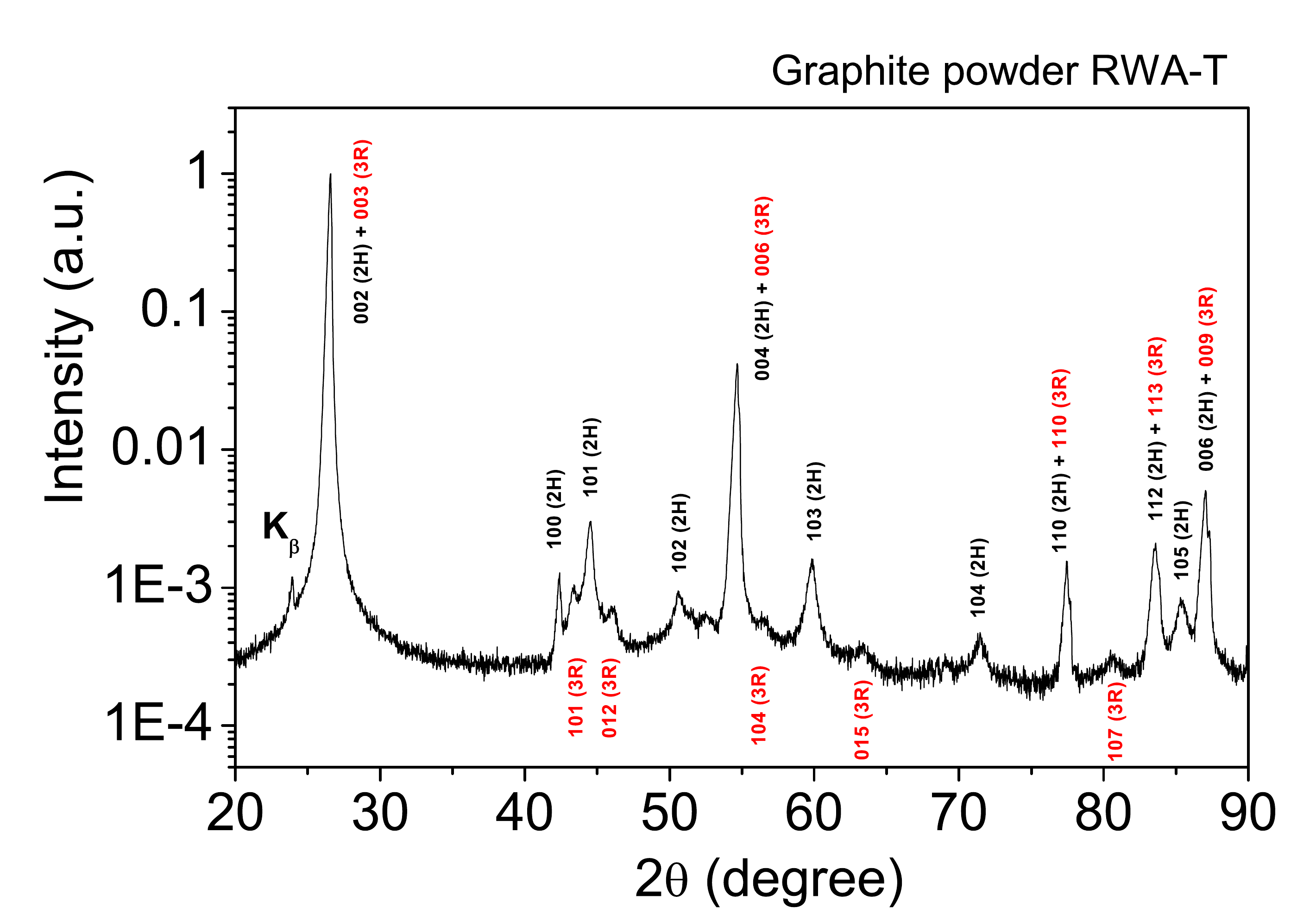}
\caption{X-rays diffraction pattern for an ultra pure graphite
powder at room temperature. The labels near the Bragg peaks
indicate whether the maximum belong to Bernal (2H) or rhombohedral
(3R) phase. Note that some of the maxima coincide with both phases
within experimental resolution, see also Table~\ref{table}.}
\label{powder}
\end{center}
\end{figure}

\begin{figure}[b!]
\includegraphics[width=1.0 \linewidth]{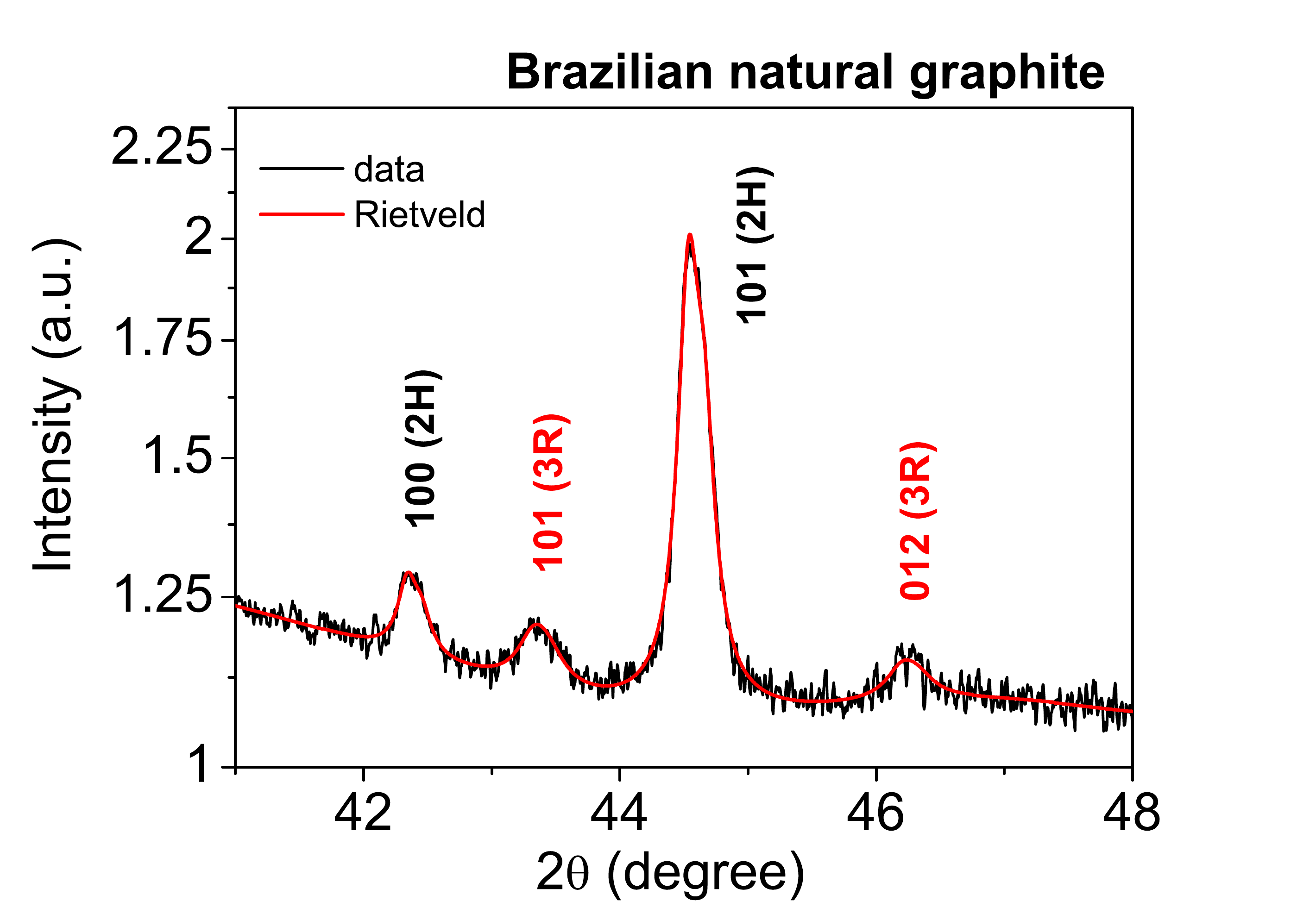}
\caption{X-rays diffraction pattern in a restricted diffraction
angle region for a natural graphite sample from Brazil. The (red)
line through the measurement is a fit used by the Rietveld
refinement to estimate the amounts of the phases.} \label{natural}
\end{figure}

The diffraction pattern of an ultra pure polycrystalline powder
RWA-T, see Fig.~\ref{powder}, shows all reflexes of the 2H and 3R
phases of graphite. This powder treated with water shows granular
superconducting-like hysteresis loops in the
magnetization\cite{sch12}. Note that there are several peaks not
suitable for distinguishing both modifications: $00l$ reflexes
with $l=2n$ for 2H and with $l=3n$ for 3R are as superposed as the
$(hh0)$ reflexes, see Table~\ref{table}. Therefore, the $2\Theta$
range $40^\circ-47^\circ$ was selected to determine and
approximately quantify the 3R phase in the samples. The selected
Bragg maxima are at $2\theta = 42.223^\circ$ and $44.393^\circ$
for the 2H phase and $2\theta = 43.451^\circ$ and $46.334^\circ$
for the 3R phase.

\begin{figure}
\includegraphics[width=1.0 \linewidth]{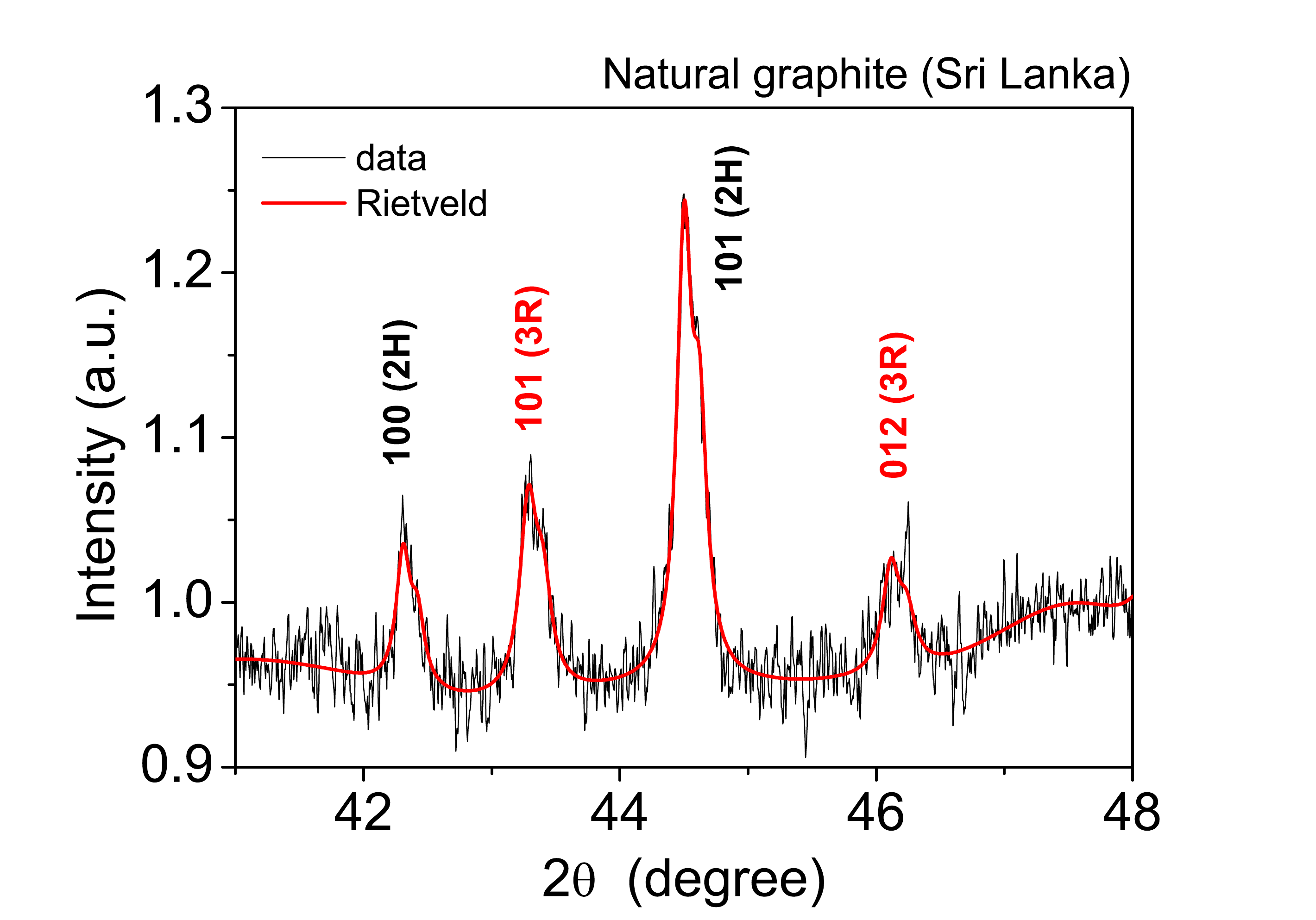}
\caption{X-rays diffraction pattern in a restricted diffraction
angle region for a natural graphite sample from Sri Lanka.}
\label{AU-xrays}
\end{figure}

\begin{figure}
\includegraphics[width=1.0 \linewidth]{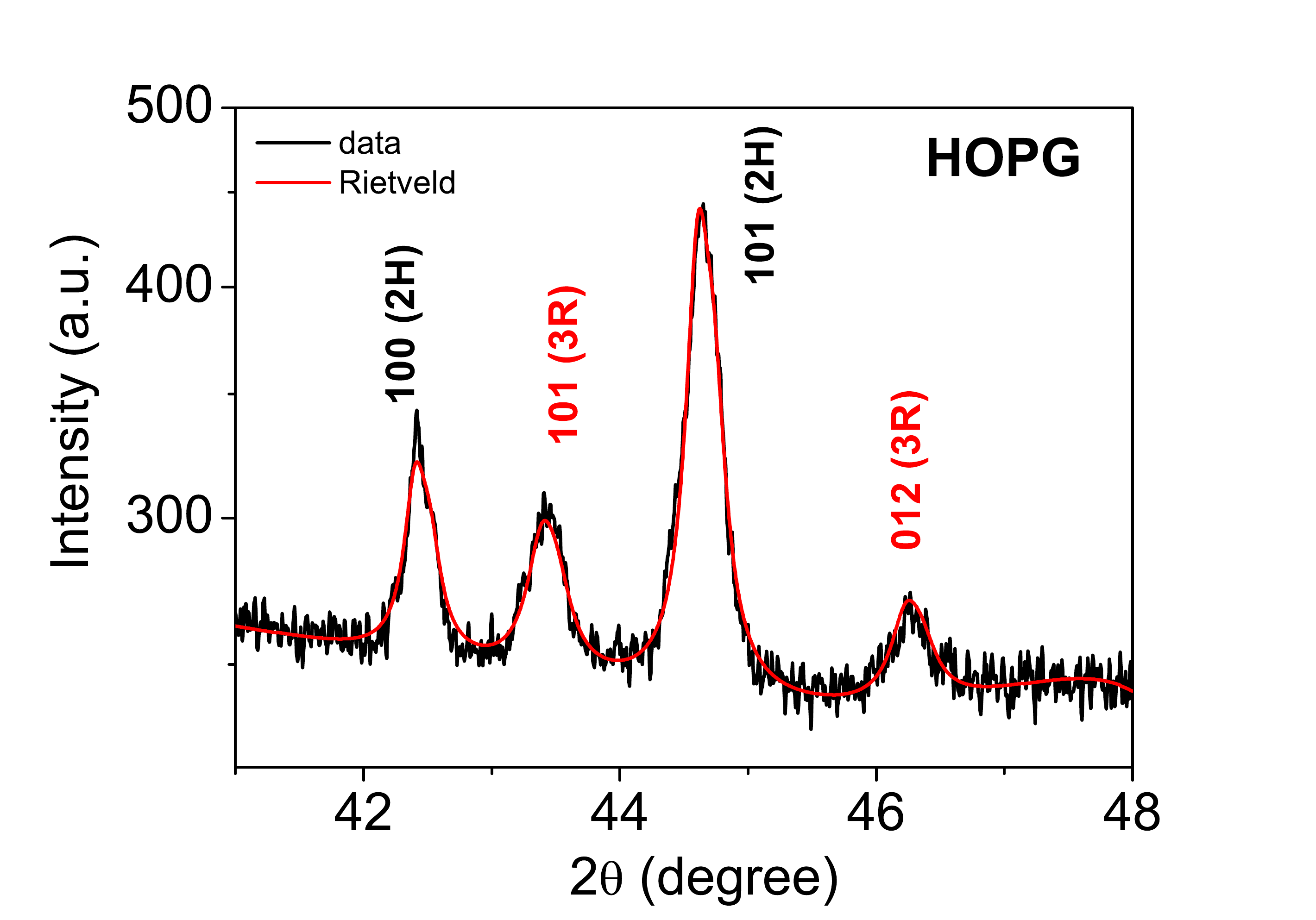}
\caption{X-rays diffraction pattern in a restricted diffraction
angle region for a HOPG sample.}  \label{hopg}
\end{figure}

In natural graphite samples from Brazil (Fig.~\ref{natural}), in
natural graphite samples from Sri Lanka (Fig.~\ref{AU-xrays}), as
well as in HOPG samples (Fig.~\ref{hopg}), the 3R phase is clearly
identified. The Rietveld refinement using TOPAS 4.2 applied on the
diffraction patterns obtained at different positions of the
natural graphite samples from Brazil gives $83 \ldots 89$~wt.\%
for the 2H and $17 \ldots 11$~wt.\% for the 3R phase.  The same
refinement applied on the patterns of several natural graphite
samples from Sri Lanka gives a percentage range for the 3R phase
between 5\% and 32\%. For the sample of Fig.~\ref{AU-xrays} the
refinement gives $75 \pm 7$~wt.\% for the 2H and $25 \pm 7$~wt.\%
for the 3R phase.

For the HOPG sample (grade A), see Fig.~\ref{hopg}, we obtained
similar quantities as for the Brazilian graphite sample within
experimental error, i.e. $\sim 86\%$ for the 2H phase and $\sim
14\%$ for the 2H phase. We note that the magnetization of a
similar HOPG sample with interfaces showed granular
superconductivity behavior (hysteresis loops, partial flux
expulsion as well as flux creep) in the magnetization at room
temperature\cite{schcar}, similar to the water-treated graphite
powder\cite{sch12}, whereas a HOPG sample without interfaces did
not\cite{schcar}.

We note that the overall x-ray diffraction characterization
presented in this study is rather unique because of the
identification of the existence of well ordered 3R phase in all
graphite samples that show hints for the existence of
high-temperature superconductivity.

\onecolumngrid

\section{Electrical resistance vs. temperature: First hint for a
transition at high temperatures}

\begin{figure}
\includegraphics[width=0.7\textwidth]{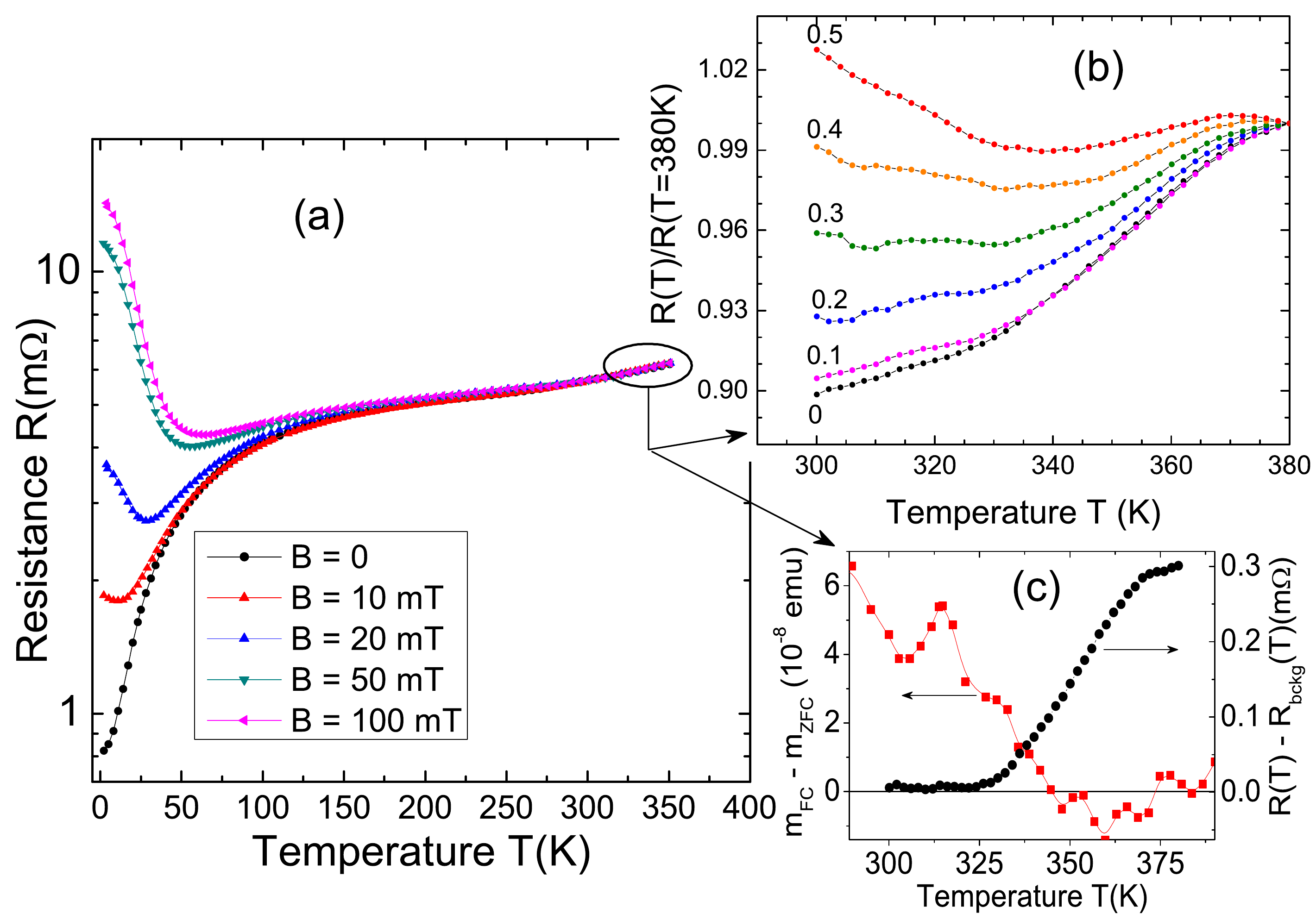}
\caption{Temperature dependence of the longitudinal electrical
resistance of a Brazilian natural graphite: (a) At different
magnetic fields applied normal to the graphene planes and
interfaces. (b) Normalized resistance data at high temperatures.
The numbers at the curves indicate the field in Tesla applied at
380~K. The measurements were done after the application of the
magnetic field and decreasing temperature. (c) Temperature
dependence of the difference between FC (field cooled) and ZFC
(zero field cooled) magnetic moments (left $y-$axis) measured with
a SQUID, after applying a field of 50~mT at 250~K with the sample
in the virgin state. The natural graphite sample is from the same
batch as the one used for the resistance measurements. Right
$y-$axis: Temperature dependence of the resistance at zero field
from (b), after subtracting a straight line background to show the
clear change in slope of $R(T)$, indicating a well-defined
transition region. The natural graphite sample had 0.4~mm length
between voltage electrodes and $\simeq 1~$mm width. } \label{mit}
\end{figure}

\twocolumngrid

Because  interfaces between the 2H and 3R phases are embedded
inside the samples, there is no simple method to make electrical
contacts directly on the interfaces. In principle, TEM lamellae
(typical size $\sim 10~\mu$m$^2$ area and $\sim 500$~nm thick)
from natural graphite samples would enable us to contact the
interfaces edges directly, as done for HOPG samples\cite{bal13}.
However, the several micrometers long interfaces in these
crystals, see Fig.~1, influence the mechanical stability of the
thin TEM lamella in such a way that its further and necessary
handling is intricate, breaking the sample before fixing and
contacting it on the substrate. Moreover, there is actually no
precise knowledge in which part of the sample and at which depth
the interfaces of interest  are localized in a sample of several
mm$^2$ area. Therefore, we decided to make the electrical contacts
at the top surface of the flakes measuring the voltage response of
an area of $\sim 0.5~$mm$^2$. In this case, it is clear that no
zero voltage is expected in case one path inside the sample gets
superconducting. Taking into account previous work\cite{esqpip},
in samples of  this size we expect to observe some behavior
compatible with the existence of  (granular) superconductivity
 at certain interfaces.

The temperature dependence of the resistance at different magnetic
fields applied normal to the interfaces and graphene planes (the
field component \cite{kempa03} that triggers the extraordinarily
large metal-insulator transition (MIT) at $T < 200~$K) is shown in
Fig.~\ref{mit}(a). A recent discussion of the origin of the MIT in
graphite can be read in \cite{esqpip}. It is related to the
existence of certain interfaces in bulk samples and it is not
intrinsic of the ideal graphite structure \cite{gar12,mah16}. The
MIT is not the main scope of this study but a transition we found
at much higher temperatures.

A careful look at the temperature dependence of the resistance at
$T > 300~$K provides the first hint of a transition at $T \simeq
350~$K, see Figs.~\ref{mit}(b) and \ref{mit}(c). The existence of
this transition is the main message of this work. Results obtained
on different samples show that this step-like behavior in the
resistance starts at similar or even higher temperatures.

\begin{figure}[]
\includegraphics[width=0.9 \linewidth]{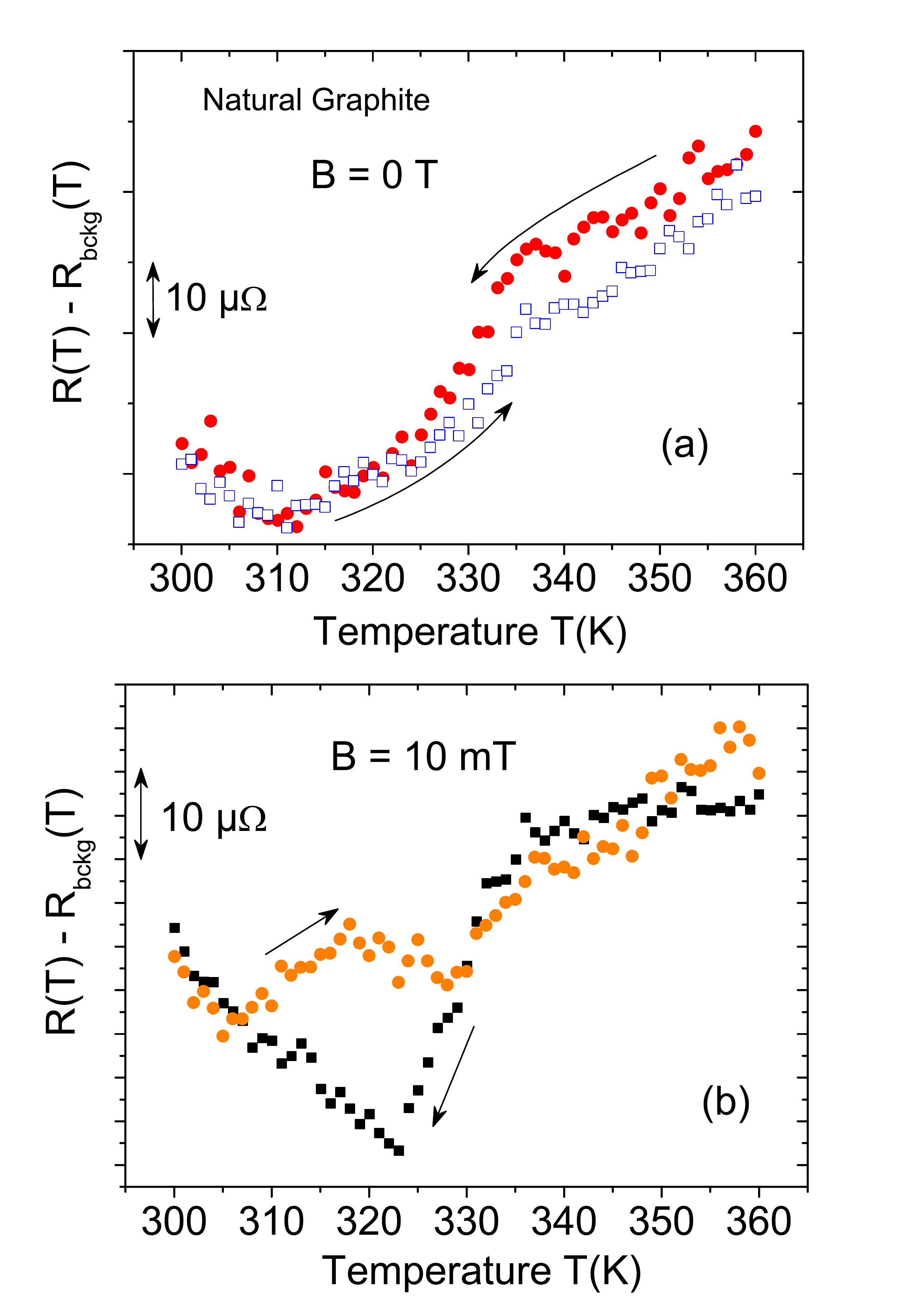}
\caption{(a) Temperature dependence of the resistance of a bulk
natural graphite sample (Brazil)  at zero field, first cooling
down and then warming up, starting at 360~K.  (b) Similar as in
(a) but after applying a magnetic field of 10~mT at 360~K. Note
that a linear in temperature background was subtracted from all
experimental curves to facilitate the recognition of  the
transition step in the resistance and the differences in the
temperature loop.} \label{IR-NG}
\end{figure}

Figure~\ref{IR-NG} shows the temperature dependence of the
resistance around the transition of a bulk natural graphite sample
(Brazil), similar to the one shown in Fig.~\ref{mit}. Each point
was obtained in thermal equilibrium after waiting 5 to 10 minutes
before measuring the signal. Afterwards we heated or cooled the
sample to the next preselected temperature. Figure~\ref{IR-NG}(a)
shows a weak temperature irreversibility at nominal zero field. We
note that the rest field of the superconducting solenoid was
measured by a Hall sensor especially designed for the measurement
of very small fields. The background field at nominally zero field
was below 0.02~mT after applying a maximum field of 10~mT. The
starting temperature was 360~K for both sets of measurements in
Figs.~\ref{IR-NG}(a) and \ref{IR-NG}(b). It is interesting to note
the difference in the irreversibility between the curves taken at
nominally zero field (a) and at 10~mT (b). The zero field
measurement shows the warming curve partially below the one
measured first by cooling, though this hysteresis is small. The
opposite behavior is observed when a low  field is applied.

\begin{figure}[]
\includegraphics[width=0.9 \linewidth]{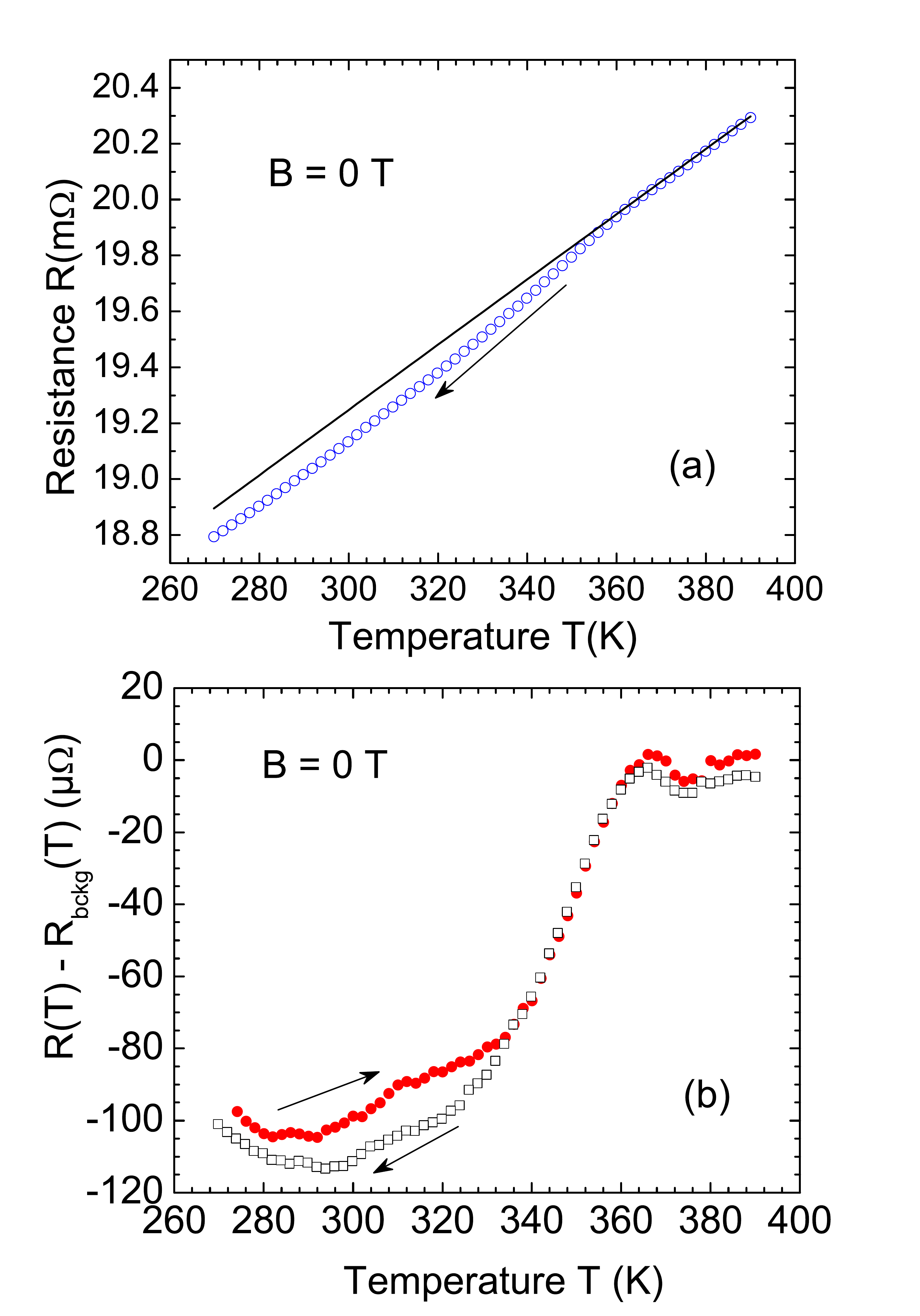}
\caption{(a) Resistance vs. temperature at zero field for a
natural graphite sample from Sri Lanka. The line is an arbitrary
line taken as background. (b) Difference between the measured
temperature dependence of the resistance and the linear dependence
taken as background, see (a). The two curves were obtained by
warming and cooling at two different days.} \label{AU-R}
\end{figure}

Figure~\ref{AU-R} shows the transition in the resistance of a bulk
natural graphite sample from Sri Lanka. In Fig.~\ref{AU-R}(a) we
show the background line subtracted to show the transitions in
(b). The two curves in (b) were obtained in two different days by
cooling and warming. The observed difference indicates the
reproducibility of our measurements. The sample was heated to
390~K several times during a whole week, before obtaining the
second (cooling down) curve.

\section{Influence of a magnetic field}

\subsection{High magnetic field influence on the high temperature
transition}

Under a magnetic field and for usual bulk superconductors we
expect a shift of the superconducting transition $T_c$ to lower
temperatures, given by the upper critical field $B_{c2}(T)$.
However, in case of  (granular) superconductivity in low
dimensional systems, as in an interface,  $B_{c2}(T)$ can be much
larger than its value in bulk and, moreover,  $T_c$ can even
increase with field\cite{gard11} or could remain field independent
as in carbon nanotubes\cite{wan12}. The transition at different
magnetic fields (applied always normal to the interfaces and
graphene planes) in field cooling mode can be seen in
Fig.~\ref{mit}(b). It is convenient to plot the normalized
resistance vs. temperature to minimize the effect of the increase
of the resistance with field. Whereas the resistance step is
clearly recognized at different fields, the transition does not
show a clear shift  to lower temperatures with field, within the
shown field range. Note that the magnetoresistance  is clearly
larger at temperatures below $T \sim 350~$K. This result also
suggests a non-simple origin of the observed transition.

\begin{figure}[]
\includegraphics[width=0.9 \linewidth]{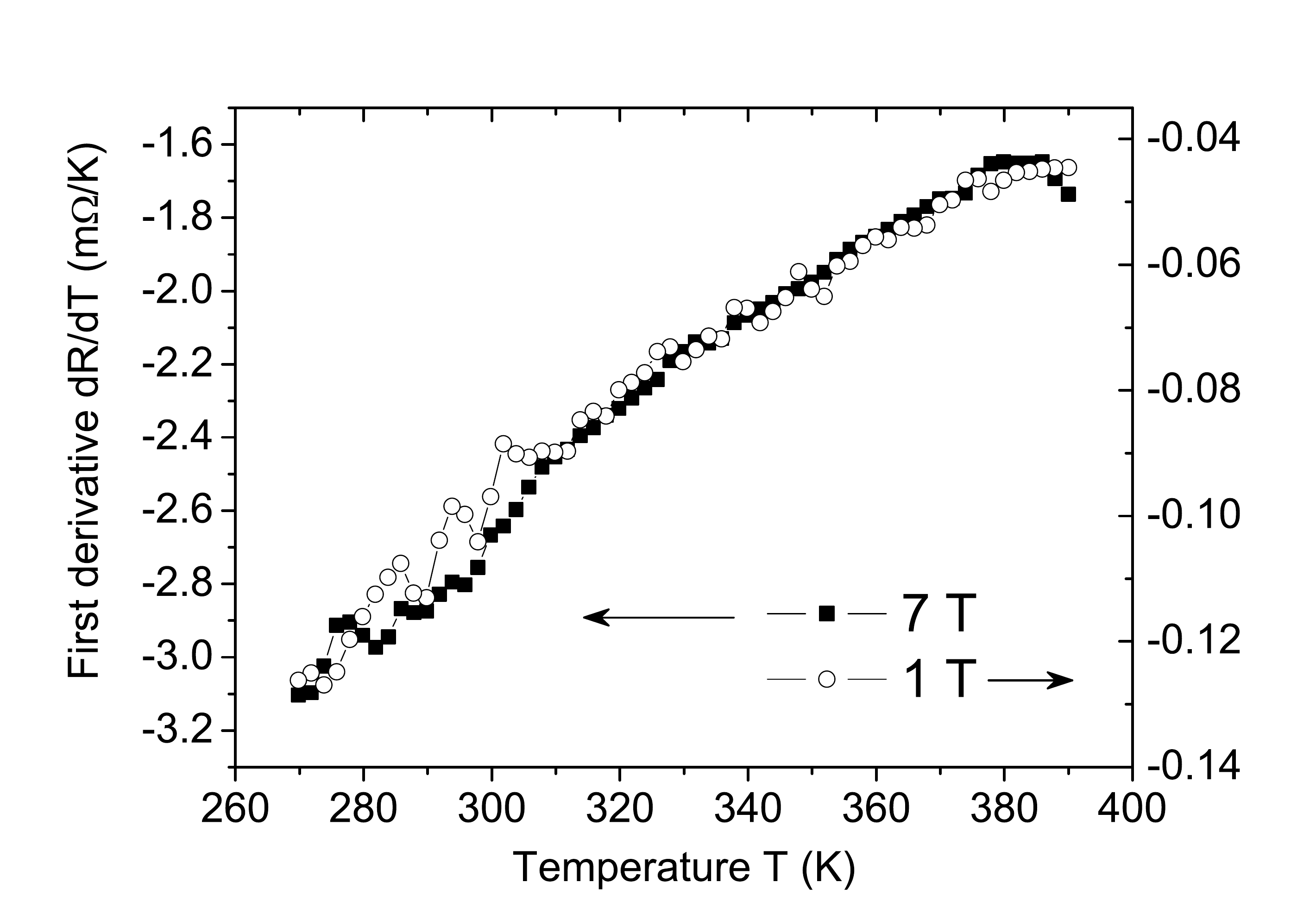}
\caption{Temperature dependence of the first temperature
derivative of the resistance at two different applied magnetic
fields for a natural graphite bulk sample from Sri Lanka.}
\label{AU-dRdT}
\end{figure}

In order to obtain some hint of the upper critical field we have
measured the temperature dependence of the resistance at different
magnetic fields. Figure~\ref{AU-dRdT} shows the temperature
derivative of the resistance at two different magnetic fields of
the bulk natural graphite crystal from Sri Lanka, see
Fig.~\ref{AU-R}. A flattening of this derivative at $T \gtrsim
375~$K is observed. The temperature at which this flattening
starts  does not show a clear shift with field within the
available field range.

\begin{figure}[b!]
\includegraphics[width=0.9\linewidth]{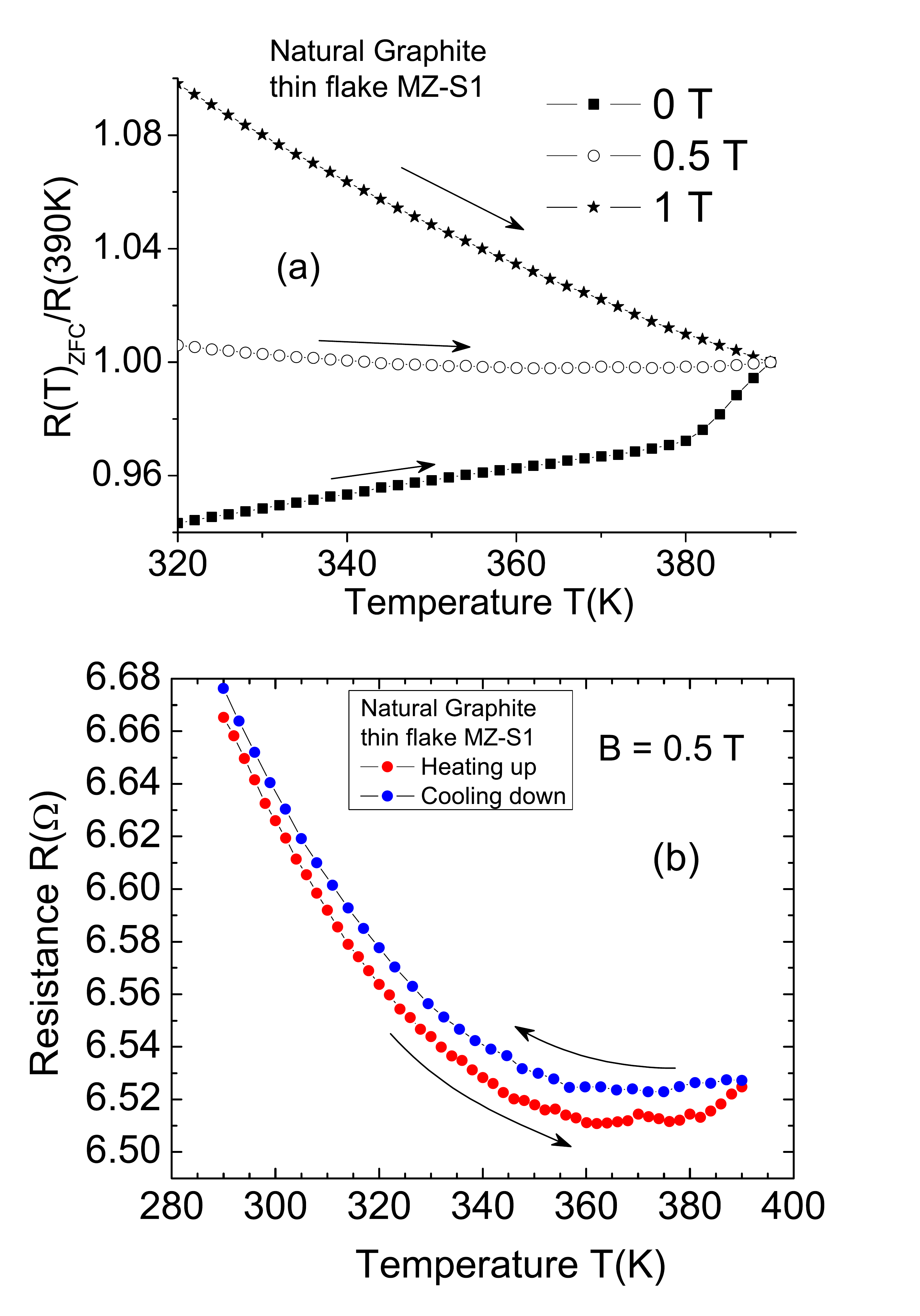}
\caption{(a) Resistance vs. temperature at three applied fields
for a thin ($\sim 200~$nm thick natural graphite flake with
voltage electrodes distance of $\simeq 5~\mu$m. All the curves
were measured after ZFC state from 390~K and applying the field at
290~K. (b) Temperature hysteresis of the resistance at a field of
0.5~T applied at 290~K. Note the slight increase in the resistance
in the heating up curve at $T \gtrsim 380~$K.} \label{NG-MZ-S1}
\end{figure}

We show results of a thin flake obtained from a bulk natural
graphite sample from Brazil. The thickness of the flake was $\sim
200$~nm and several micrometers length and width. The Pd/Au
electrical contacts were made using electron lithography, see e.g.
Ref.~\onlinecite{esq08}. Figure~\ref{NG-MZ-S1}(a) shows the
electrical resistance measured after  zero field cooling (ZFC)
from 390~K to 290~K  and measured by warming after applying the
magnetic field at 290~K. After reaching 390~K, the sample was
cooled down at zero field again and the next measurement at higher
field was done. At all applied fields we found a small temperature
hysteresis, see Fig.~\ref{NG-MZ-S1}(b).

Figure~\ref{NG-MZ-S1}(b) shows the resistance at a field of 0.5~T
applied in the ZFC state at 290~K. This measurement was done after
the zero field measurement shown in (a). The origin of this
hysteresis can be due to: (a) an irreversible change in the
resistance due to an annealing process at the relatively high
temperatures (and measuring time) and/or (b) flux pinning. Because
the flux free virgin state cannot be fully reached if the onset of
the transition is above 390~K, as it appears for this sample (see
the curve at zero field in Fig.~\ref{NG-MZ-S1}(a)), it is not
possible to discern, which of the two possibilities is the origin
of the irreversibility, unless one takes the sample out of the
cryostat and anneal it above 390~K in an independent oven. For
that, however, other technical problems appear especially with the
electrical contacts and the already fixed wires.  As in all the
other temperature dependent measurements, each point was obtained
after waiting for thermal equilibrium (no temperature sweeping
mode), i.e. each point took about 5 to 10 minutes waiting time.

We may conclude that the large magnetoresistance makes difficult a
clear determination of $B_{c2}(T)$, in case it exists, at this
stage and for the usual electrode configuration. Nevertheless and
within the obtained data it appears that the $B_{c2}(T)$ is
significantly larger than the one measured in the CuO$_2$-based
high-temperature superconductors. The magnetization measurements
shown below support this conclusion. Taking into account the
two-dimensionality of the superconducting regions localized at the
interfaces, a high bulk critical field is in principle expected,
see  Section~\ref{dis}.

\onecolumngrid

\subsection{Low magnetic field response: irreversibility and remanence}

\begin{figure}
\includegraphics[width=0.85\textwidth]{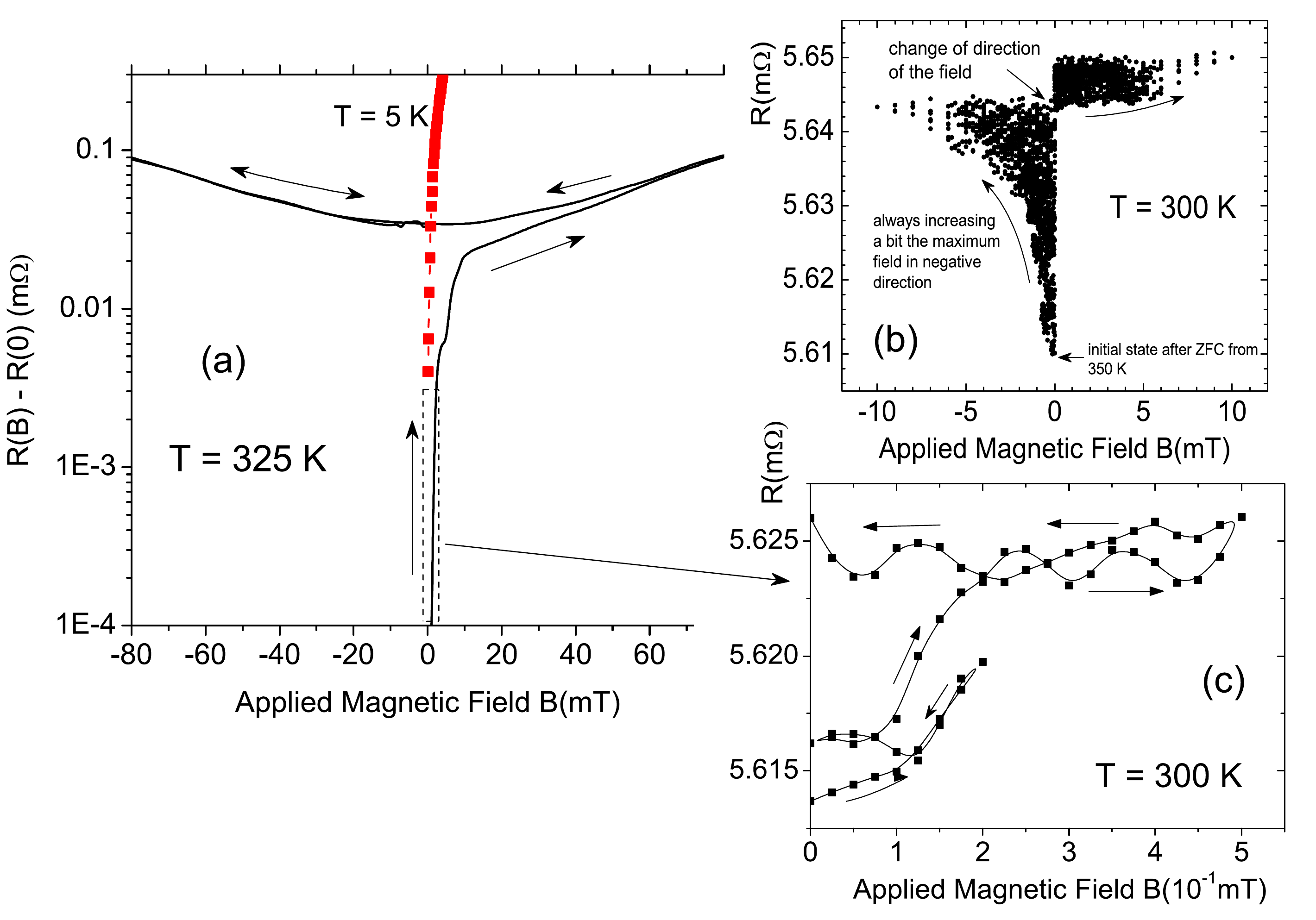}
\caption{Magnetoresistance at low fields of a Brazilian natural
graphite: (a) Difference between the resistance and its value at
zero field vs. applied field in two opposite directions at 325~K.
The arrows indicate the sweep field direction. The red points were
measured at 5~K. (b) Similar as in (a) but at 300~K and increasing
stepwise the maximum field after returning to zero field. After
reaching $-10$~mT the field direction is changed to the opposite
direction starting with a maximum field of 5~mT. (c) Very low
field response of the sample starting in its virgin state at 300~K
after ZFC from 390~K.} \label{ir}
\end{figure}

\twocolumngrid

Taking into account the low-field sensitivity of earlier transport
measurements in annealed graphite powder samples \cite{ant74} and
the hints for the existence of the Josephson-effect in HOPG
lamellae \cite{bal13},  in this work we have studied carefully and
with high resolution the change of the electrical resistance with
field, in the low field region. Figure~\ref{ir}(a) shows the
magnetoresistance at 325~K of a bulk natural graphite sample from
 Brazil, starting after ZFC from 390~K. The measured virgin
 curve starting at zero field shows a rapid increase in
 the resistance with field. At around 10~mT it shows a clear
decrease in the slope and continues increasing slowly with field.
Note that after applying a field $\sim 60~$mT or higher, the
magnetoresistance shows a quadratic in field and reversible
behavior, within experimental error, see Fig.~\ref{ir}.

We note that graphite flakes shows an anomalous field hysteresis
but at $T < 25~$K compatible with the existence of Josephson
coupled superconducting regions \cite{esq08}. Due to the
resolution limits and the relatively large background, for the
observation of the small anomalous field hysteresis it is
necessary to restrict the sample size by, for example, introducing
micro-constrictions \cite{sru11}. The irreversibility we find at
much higher temperatures in the studied samples and discussed in
this work does not show a field hysteresis but has a different
character and it is observed in large samples, although it can be
also related to granular superconductivity, as we explain below.

Figure~\ref{ir}(c) shows the change in resistance and its
irreversible behavior at fields below 0.5~mT at 300~K, starting
from the ZFC state. Figure 3(b) shows similar data but increasing
a small amount the maximum applied field after each field cycle.
At a given resistance/field value we changed the direction of the
field to show that the resistance does not return to its virgin
state. It is important to note that to return to the virgin state
one needs to heat the sample at $T \gtrsim 350~$K and cool it at
zero field. This remanence in the resistance after applying such
small fields indicates the existence of trapped magnetic flux.

\begin{figure}[]
\includegraphics[width=1\linewidth]{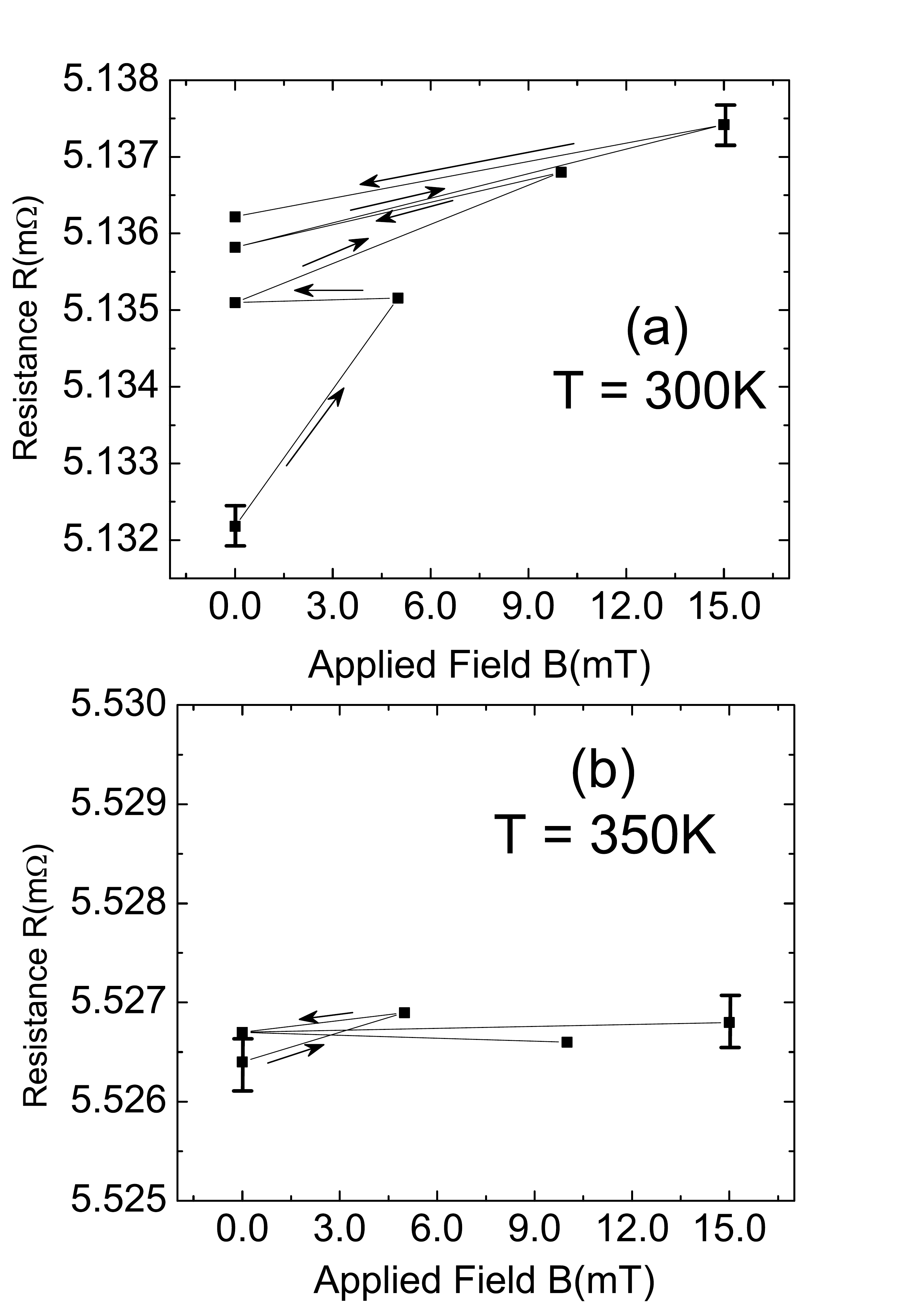}
\caption{(a) Resistance vs. applied field applied following the
arrows for the same natural graphite bulk sample (Brazil) as shown
in Fig.~\ref{mit}. The initial state at zero field was reached
after cooling the sample from 390~K at zero field to the
respective temperature. (b) The same as in (a) but at 350~K, i.e.
at the middle of the transition in the resistance, see
Fig.~\ref{mit}(b,c).} \label{IR-300}
\end{figure}

Figure~\ref{IR-300} shows the irreversibility in the resistance at
a temperature below the transition (300~K, Fig.~\ref{IR-300}(a))
and in the transition region (350~K, Fig.~\ref{IR-300}(b)), for a
bulk natural graphite sample similar to the one shown in
Fig.~\ref{mit}. The starting zero-field cooled state was reached
after cooling the sample from 390~K. Note the absence of remanence
at 350~K within experimental error. This result is a further
proof for the existence of a transition at $\sim 350~$K, as shown
in Fig.~\ref{mit}.

\begin{figure}
\includegraphics[width=0.5\textwidth]{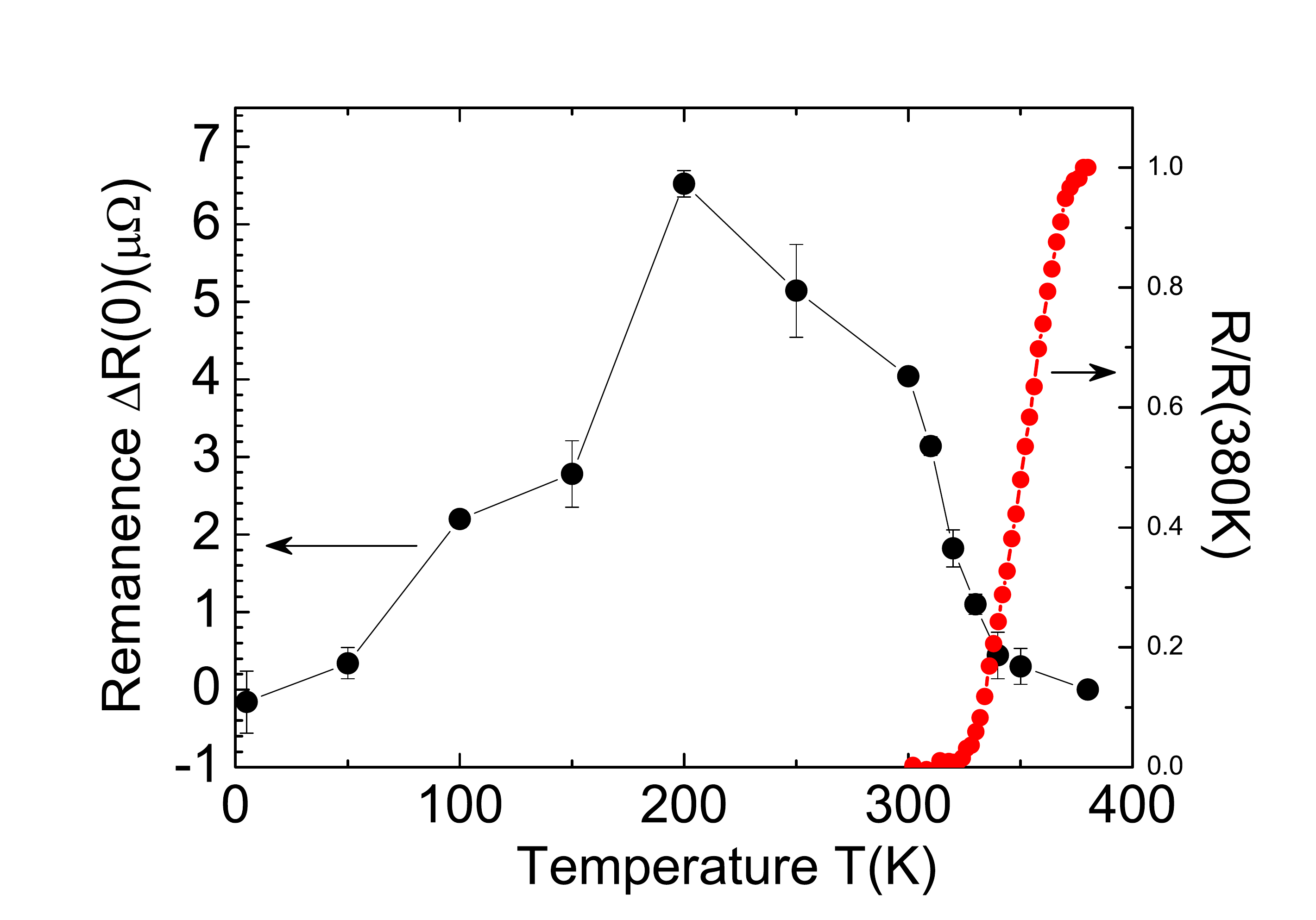}
\caption{Temperature dependence of the remanence of a Brazilian
natural graphite crystal. Each point was measured after ZFC from
390~K. $\Delta R(0)$ is the difference between the resistance at
zero field after applying a field of 10~mT and the resistance at
zero field in the virgin state, prior to the application of 10~mT.
The red points (right $y-$axis) represent the resistance
normalized by the resistance at 380~K at zero field, after
subtraction of a linear in temperature background (a straight
line), similar to Fig.~\ref{mit}(c).} \label{rem}
\end{figure}

To quantify the temperature dependence of the remanence we used a
three-points method that consists in: 1) cooling the sample at
zero field from 390~K to the preselected temperature, 2) measuring
the resistance at zero field $R_0(0)$, 3) applying a certain
magnetic field, 4) returning to zero field and measure the
resistance again. The remanence is then defined as the difference
between these two resistances at zero field $\Delta R(0)$.
Figure~\ref{rem} shows this difference vs. temperature. The
transition at $T \sim 350~$K is clearly discernible in the
remanence and it shows a maximum at $T \sim 200~$K, tending to
zero at the lowest temperature.

\begin{center}
\begin{figure}[]
\includegraphics[width=0.9\linewidth]{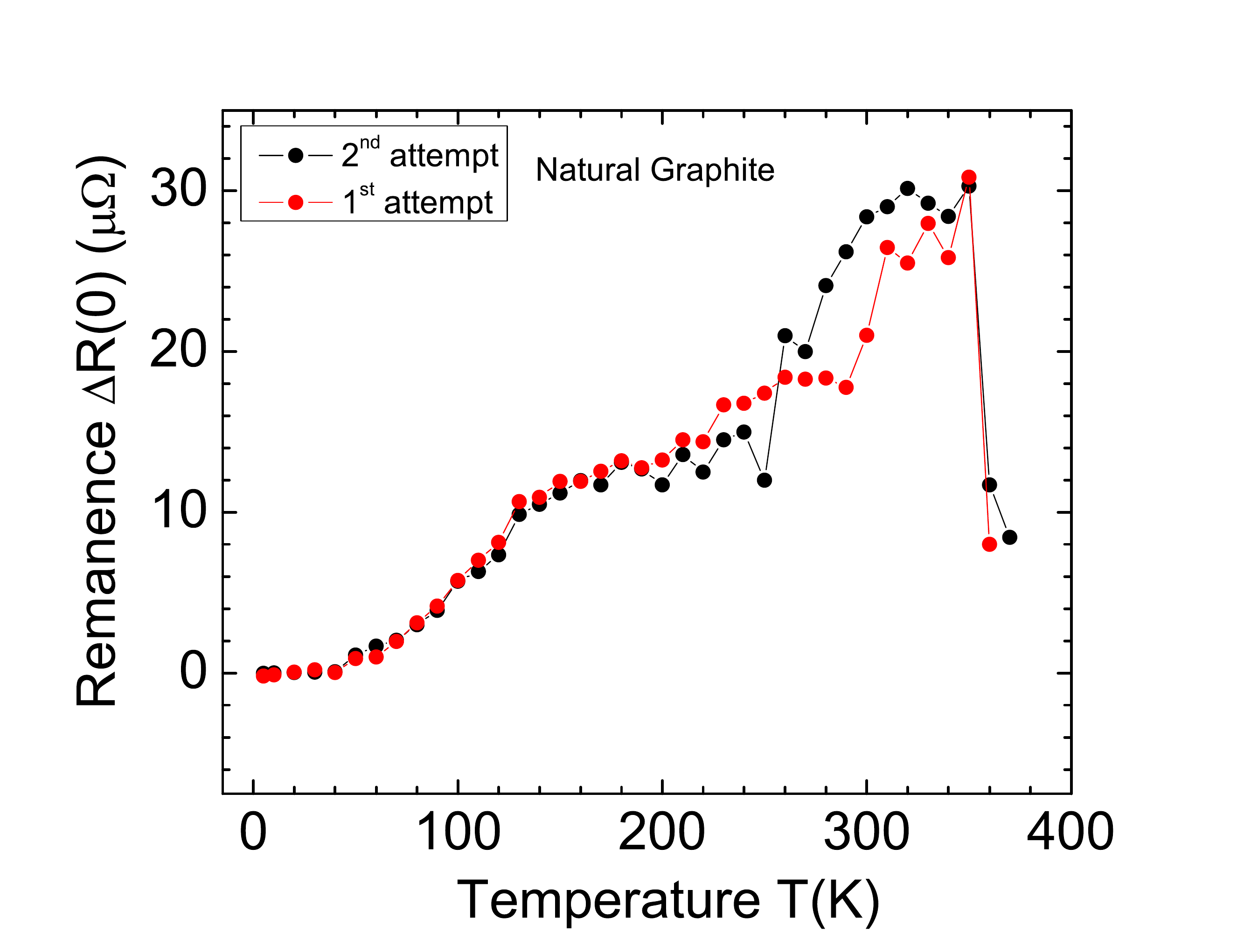}
\caption{Remanence defined as $\Delta R(0) = R_{10}(0) - R_0(0)$,
where $R_{10}(0)$ is the resistance measured at zero field after
applying a field of 10~mT, and $R_{0}(0)$ means the resistance at
zero field in the virgin state after ZFC from 390~K. The sample
was a natural graphite bulk sample (Brazil) different from the
sample shown in Figs.~\ref{IR-300} and \ref{rem}.} \label{RE-2}
\end{figure}
\end{center}

Figure~\ref{RE-2} shows the remanence of a different bulk natural
graphite sample obtained using the three-points method. The two
curves shown in the figure were obtained in the same way but after
waiting for a couple of days between the two measurements. Note
that each point was obtained starting with the sample in the
virgin state, as the results of Fig.~\ref{rem}. It means that
before magnetizing the sample at a given temperature, it was
heated to 390~K and cooled down at the respective temperature at
zero field.

\begin{center}
\begin{figure}[]
\includegraphics[width=0.9\linewidth]{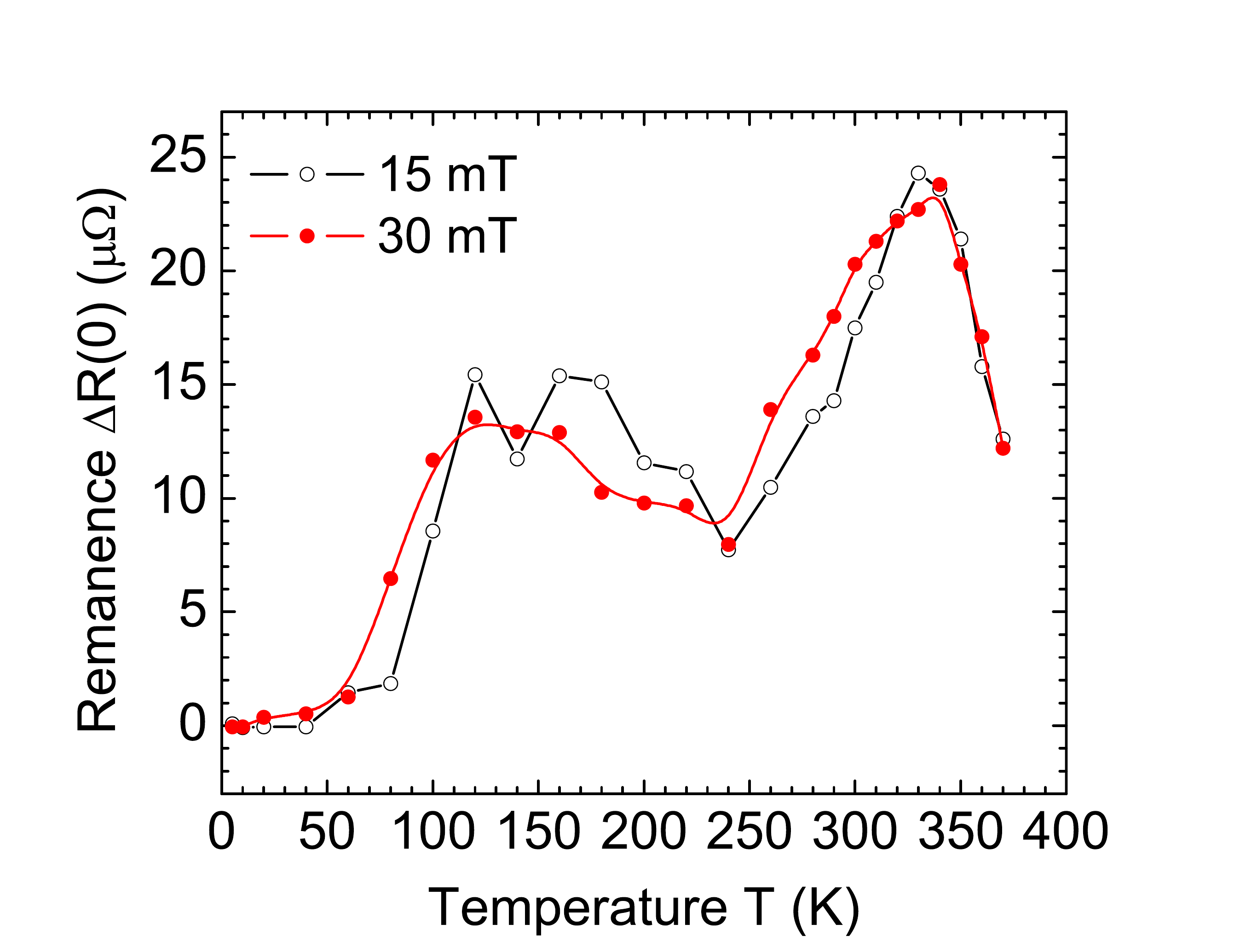}
\caption{Remanence defined as $\Delta R(0) = R_{B}(0) - R_0(0)$,
where $R_{B}(0)$ is the resistance measured at zero field after
applying a field of $B = 15$~mT ($\circ$) and 30~mT (red
$\bullet$). $R_{0}(0)$ means the resistance at zero field in the
virgin state after ZFC from 390~K. The sample was a natural
graphite bulk sample (Sri Lanka).} \label{AU-Rem}
\end{figure}
\end{center}

Figure~\ref{AU-Rem} shows the temperature dependence of the
remanence at two different applied fields for a bulk natural
graphite sample from Sri Lanka. Basically the same results are
obtained for both relatively large fields, as expected taking into
account the behavior of the magnetoresistance shown in
Fig.~\ref{ir}.

\begin{figure}[]
\begin{center}
\includegraphics[width=1.0 \linewidth]{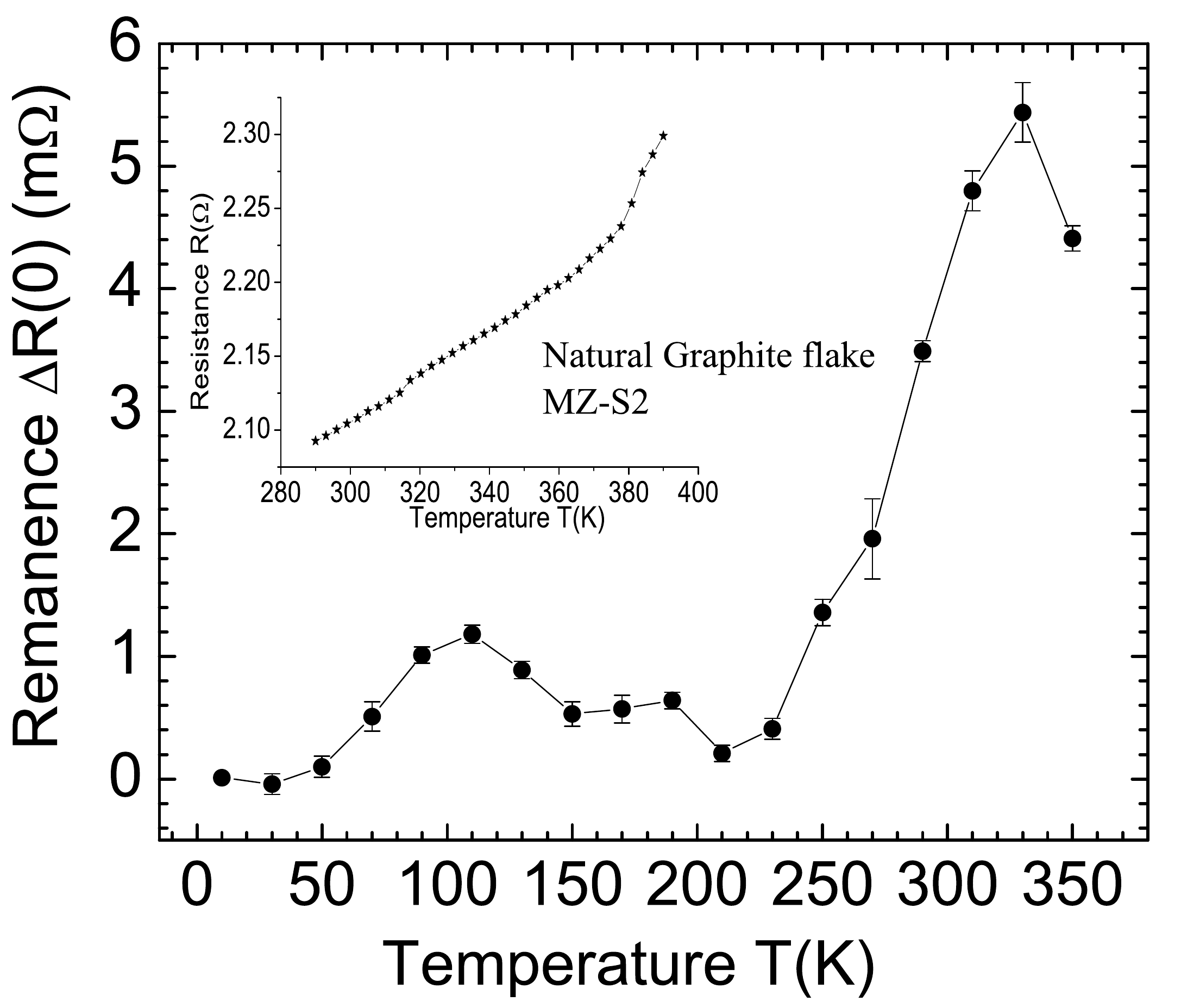}
\caption{Main panel: Remanence vs. temperature (similar to
Figs.~\ref{rem}, \ref{RE-2} and \ref{AU-Rem}) for a natural
graphite thin (thickness $\simeq 200~$nm) flake. The remanence was
defined as $\Delta R(0) = R_{15}(0) - R_0(0)$, where $R_{15}(0)$
is the resistance measured at zero field after applying a field of
15~mT, and $R_{0}(0)$ means the resistance in the virgin state
after ZFC from 390~K, at zero field. The inset shows the virgin
resistance vs. temperature at zero field measured by warming.}
\label{RE-S2}
\end{center}
\end{figure}

Figure~\ref{RE-S2} shows a similar plot  for a  natural graphite
thin flake obtained from a bulk natural crystal from Brazil.
Qualitatively, the measured temperature dependence of the
remanence resembles that obtained for bulk natural graphite
samples. The insert in that figure shows the temperature
dependence of the resistance measured increasing temperature  at
zero field.

\begin{figure}[!]
\includegraphics[width=0.9\linewidth]{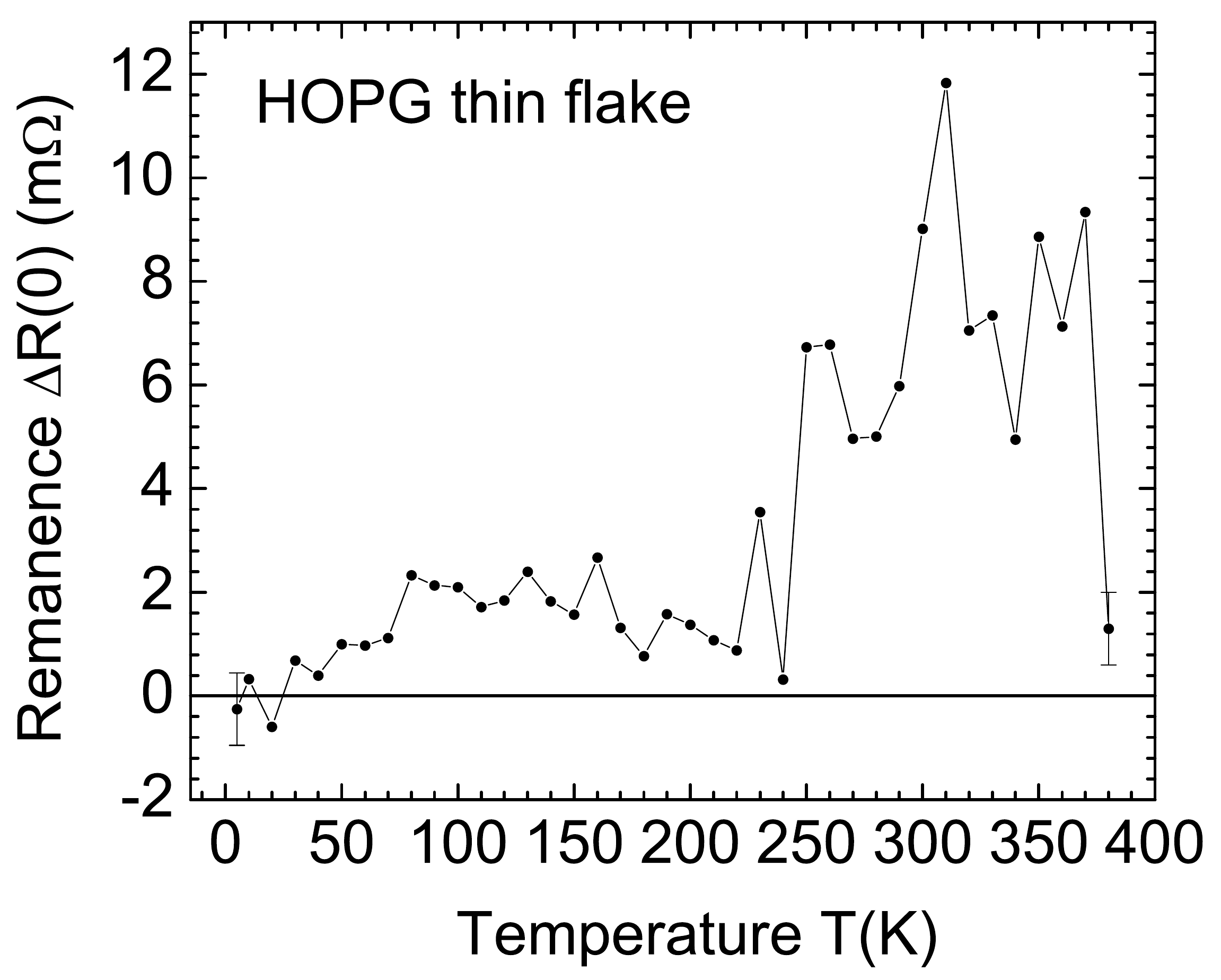}
\caption{Remanence vs. temperature for the HOPG thin flake.
The remanence was defined as $\Delta R(0) =
R_{10}(0) - R_0(0)$, where $R_{10}(0)$ is the resistance measured
at zero field after applying a field of 10~mT, and $R_{0}(0)$
means the resistance at zero field and in the virgin state after
ZFC from 390~K.}\label{rem-hopg}
\end{figure}

Similar measurements were done for a thin ($\sim 50~$nm) HOPG
flake  that has interfaces according to STEM studies, see
\cite{bar08,gar12}. Figure~\ref{rem-hopg} shows the remanence
resistance $\Delta R(0)$ vs. temperature for the HOPG flake. Note
that the temperature dependence is similar to the other measured
samples (see Figs.~\ref{RE-2}, \ref{AU-Rem} and \ref{RE-S2}) but
it has a larger scatter. This scatter is possibly due to the
higher transition temperature, i.e. with an onset at $T \gtrsim
$390~K, and a large transition width that prevent to reach always
the same flux-free virgin state before cooling to the selected
temperature.

The observed behavior for the temperature dependence of the
remanence is general for all samples we measured but the details
are sample dependent. It is interesting to note that the
irreversibility of the magnetization (difference between FC and
ZFC curves) of graphite bulk HOPG samples (with interfaces) shows
local maxima at $\simeq 305~$K, $\simeq 200~$K and $\sim 100
\ldots 130~$K, which remain independent of the applied field to
7~T,\cite{schcar} indicating a general behavior in the pinning of
trapped flux in this material.

\section{Magnetization measurements}
\subsection{Temperature dependence}

If superconductivity is localized at the interfaces and the field
is applied normal to them, the large demagnetization factor would
prevent a full flux expulsion (the Meissner effect). However,
partial flux expulsion is always possible to observe if vortex (or
fluxon) pinning exists in the superconducting sample. In this case
a positive difference between the FC and ZFC magnetic moment is
expected below $T_c$ and after applying the field at the lowest
temperature of the temperature loop cycle with the sample in the
ZFC state. Figure~\ref{mit}(c) shows the difference in the
magnetic moment of other natural graphite sample from Brazil, at
an applied field of 50~mT. Lowering temperature, this difference
starts to increase just below the transition temperature measured
with the resistance, right $y$-axis in Fig.~\ref{mit}(c).

\begin{figure}[]
\includegraphics[width=1 \linewidth]{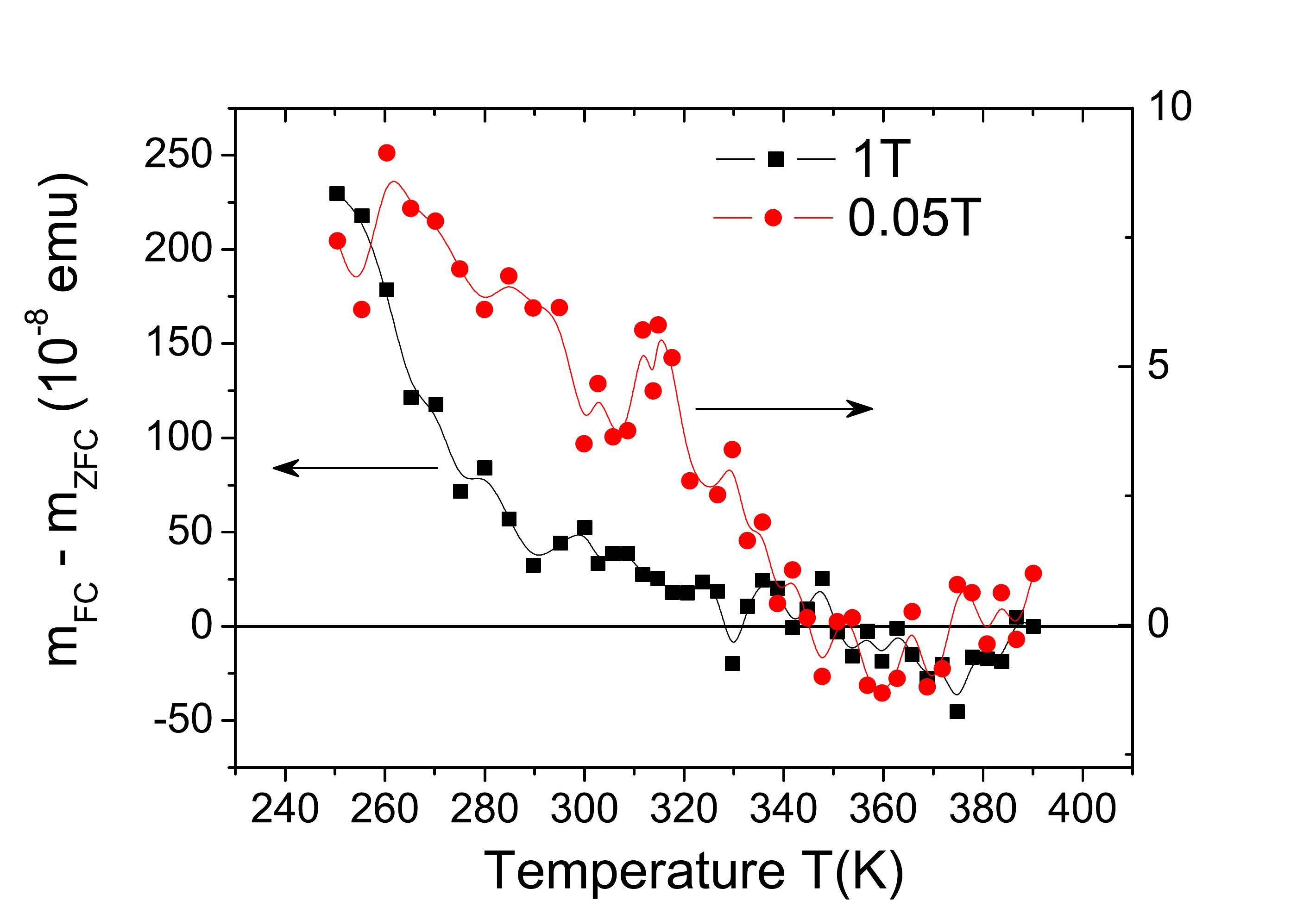}
\caption{Temperature dependence of the difference between FC and
ZFC magnetic moment at two different applied fields of a natural
graphite bulk sample (Brazil).} \label{Dm}
\end{figure}

\begin{figure}[]
\includegraphics[width=1 \linewidth]{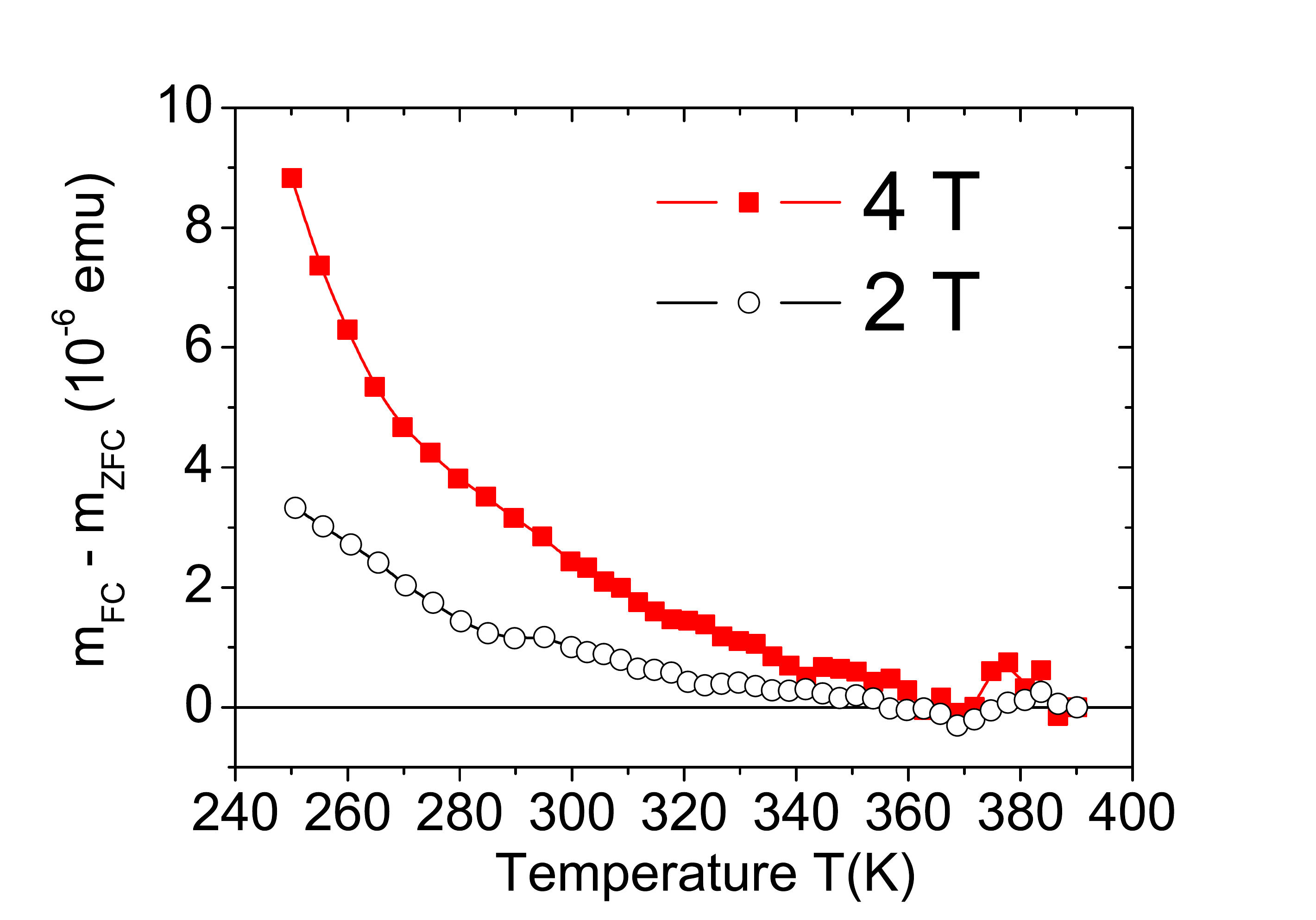}
\caption{As in Fig.~\ref{Dm} but at higher applied fields
 of the same natural
graphite bulk sample (Brazil).} \label{Dm2}
\end{figure}

Figure~\ref{Dm} shows the temperature dependence of the difference
between FC and ZFC magnetic moment at two different applied fields
of the same natural graphite bulk sample (Brazil). Note the
increase in the irreversibility values with field. In spite of
that, a clear shift of the temperature onset, where
$m_{FC}-m_{ZFC}$ starts to be larger than zero, to lower
temperatures by increasing the applied field cannot be clearly
recognized from the measurements. Measurements done of this
difference at higher fields are shown in Fig.~\ref{Dm2}.

At such high temperatures one expects that the strength and
temperature dependence of the measured irreversibility in the
magnetization is mainly due the pinning of the involved magnetic
entities (vortices and/or fluxons, see Sec.~\ref{dis}). The shift
to low temperatures with applied field of the onset in the
magnetic moment difference would be given by the depinning or
irreversibility line, as in the oxides high-temperature
superconductors. The apparently small, if at all, shift of the
onset with field in the measured field region would indicate
therefore a surprisingly large pinning strength.

\begin{figure}[]
\begin{center}
\includegraphics[width=1\linewidth]{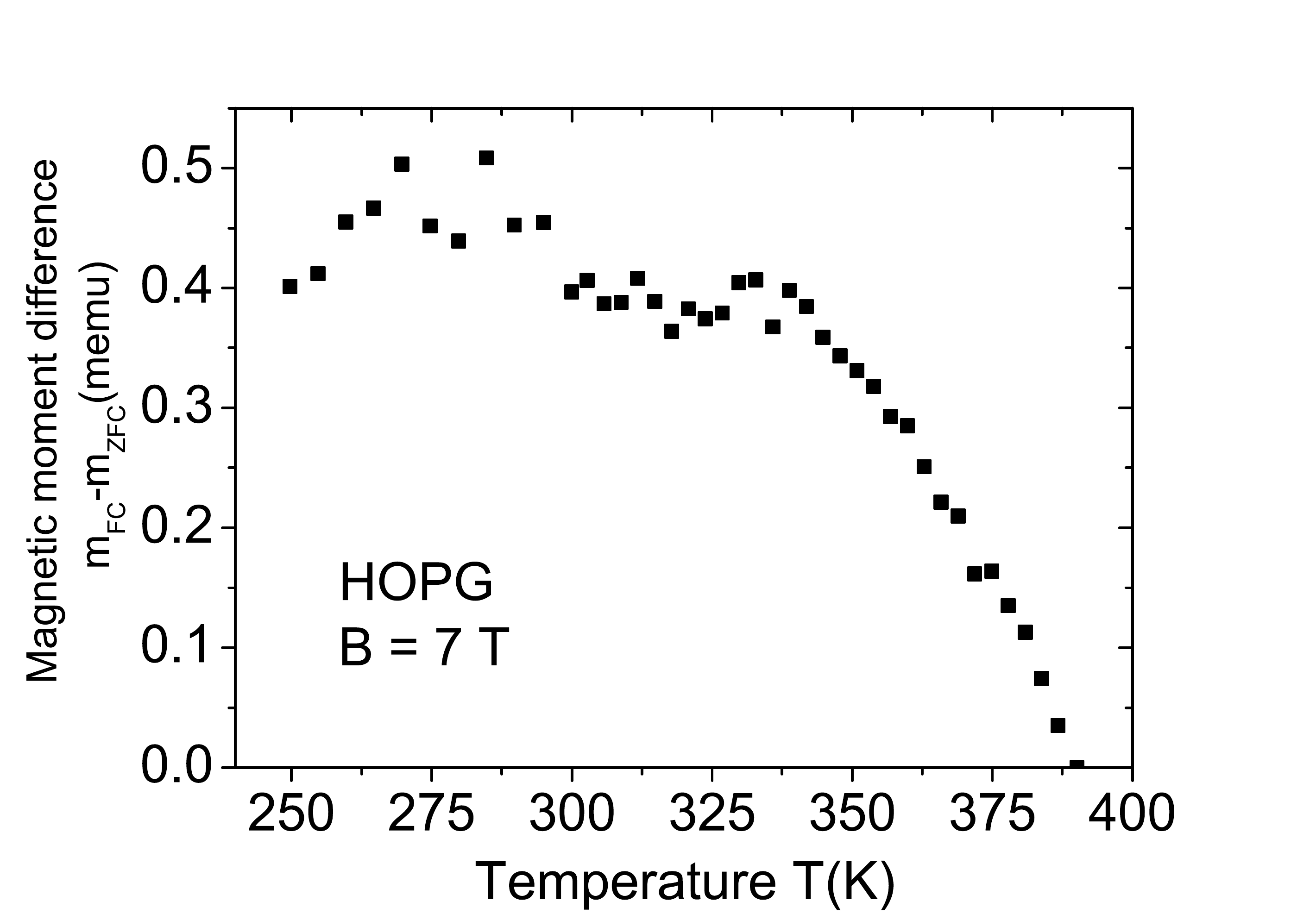}
\caption{Difference between FC and ZFC magnetic moment vs.
temperature obtained for a HOPG bulk sample at an applied field of
7T.}\label{HOPG-Dm}
\end{center}
\end{figure}

Figure~\ref{HOPG-Dm} shows the difference between FC and ZFC
magnetic moment vs. temperature at an applied field of 7T of a
bulk HOPG sample, similar to those shown in Ref.~\cite{schcar}.
The field was applied at 250~K with the sample in the virgin state
(cooled from 390~K at zero field). In agreement with previous
measurements\cite{schcar}, this difference indicates that the
onset of the (possible superconducting) transition temperature
should be above 390~K. The turning point in these measurements was
390~K.

\subsection{Time dependence}

\begin{figure}[]
\includegraphics[width=0.9 \linewidth]{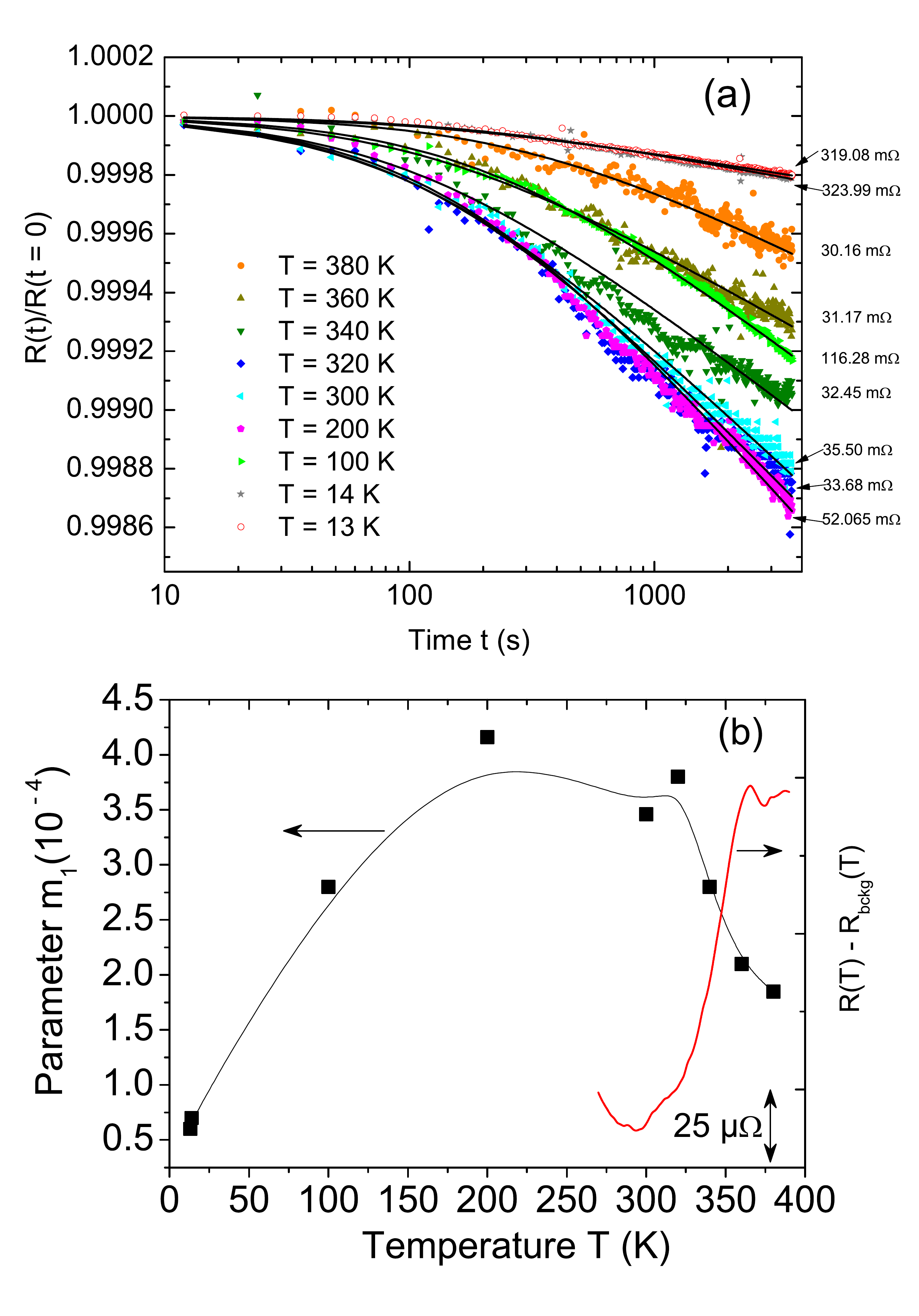}
\caption{(a) Normalized resistance vs. time at different constant
temperatures, after applying a field of 1~T from zero field
(virgin state) reached after ZFC from 390~K. The continuous lines
are fits to the function $R(t)/R(0) = 1 - m_1 \ln(1 + (t/\tau))$.
The numbers at right indicate the resistance at $t = 60~$s at the
corresponding temperature. (b) The fit parameter $m_1$ vs.
temperature (left $y-$axis) obtained from the curves in (a). The
line is only a guide to the eye. Right $y-$axis: The line is the
measured resistance transition after subtracting a linear in
temperature background, see Fig.~\ref{AU-R}(b).} \label{AU-time}
\end{figure}

Usually, due to the finite pinning strength of the magnetic
entities, one expects to measure a time relaxation in certain
properties like the magnetization or magnetoresistance. The time
dependence of the magnetization of water treated graphite
powder\cite{sch12} at zero field, after applying a finite field,
has been shown to follow the well known logarithmic  time
dependence $m(t)/m(0) = 1 - m_1 \ln(1 + (t/\tau))$ with
coefficient $m_1 \sim k_BT/U_a$, where $U_a$ is an apparent flux
creep activation energy and $\tau$ a time constant that determines
a transient stage before the beginning of the logarithmic
relaxation\cite{gur93}. Due to the small mass of the natural
graphite samples, magnetization measurements are not enough
sensitive to observe the time dependence due to a possible flux
creep. However, our resistance measurements have enough resolution
to measure this time relaxation. Figure~\ref{AU-time}(a) shows the
time relaxation for the same bulk natural graphite sample (Sri
Lanka) shown in Fig.~\ref{AU-R}. The curves were obtained applying
(at constant temperature, waiting half an hour for temperature
stabilization) a field of 1~T from the ZFC state of the sample
(starting always at 390~K). The zero time was defined 10~s after
the superconducting solenoid was set in permanent modus. The
resistance follows a similar logarithmic time dependence as the
magnetization\cite{sch12}. Note that after applying the field the
magnetoresistance decreases with time because of the entrance of
flux inside the sample, which implies a decrease of the effective
field seeing by the sample.

From the fits to the time dependence shown in
Fig.~\ref{AU-time}(a) we obtain the parameter $m_1(T)$.
Figure~\ref{AU-time}(b) shows the temperature dependence of this
parameter. It shows a clear increase decreasing temperature in the
region of the transition ($y-$axis in Fig.~\ref{AU-time}(b)). Note
that one expects that $m_1 \propto T$. This
 dependence appears to be followed below 200~K, see Fig.~\ref{AU-time}(b).

\section{Discussion}
\label{dis}

The main reasons to identify the small step in the resistance with
temperature, see Figs.~\ref{mit} and \ref{AU-R}, as a
superconducting transition are: (1) Step-like decrease in the
resistance decreasing temperature (see Fig.~\ref{mit}(b)), (2)
relatively large {\em positive} magnetoresistance at very small
applied fields and temperatures (see Fig.~\ref{ir}), (3) field
irreversibility with finite remanence at  small applied fields
(see Figs.~\ref{ir}, \ref{rem}-\ref{rem-hopg}), (4) time
dependence of the magnetoresistance after changing the apply
magnetic field, compatible with flux creep (see
Fig.~\ref{AU-time}), and (5) partial magnetic flux expulsion below
the transition (see Figs.~\ref{mit}(c), \ref{Dm}, \ref{Dm2} and
\ref{HOPG-Dm}).

In general, we can imagine two possibilities that a material shows
a magnetic field irreversibility and finite remanence, namely: a)
due to vortex pinning or flux trapping, requiring the existence of
superconducting currents, or b) magnetic order, due to the pinning
of magnetic domains or magnetic field anisotropy, that appear
below the corresponding Curie temperature and without applying any
external field. We do not expect to have magnetic order due to
impurities or defects in all graphite samples we measured because
the measured concentration of magnetic impurities is far too small
(see Sec.~\ref{details}).  Moreover, the observed
 behavior does not seem to be compatible to phenomena related
 to magnetic order in graphite, namely: - we observe a positive
instead of a negative magnetoresistance measured for magnetic
graphite\cite{jems08}, - relatively large response of
 the resistance in the virgin state of
the sample in a field range below a few mT and at room
temperature, - the virgin state resistance is not reached anymore
after applying a finite magnetic field, even changing the field
direction, - the field irreversibility and the remanence tend to
vanish at low temperatures. Magnetic field-induced structural
changes in the graphite samples at such small field strengths can
be disregarded. Note also that the transition does not change
basically after applying a magnetic field below 1~T, see
Fig.\ref{mit}(b).

The overall behavior shown in Fig.~\ref{ir} is compatible with
trapped magnetic flux in granular superconductors\cite{ji93} and
in Josephson junctions\cite{wat07}. The trapped flux can occur
within the superconducting regions, i.e. through the existence of
pinned intragranular vortices, but also within superconducting
loops enclosing regions where the Josephson junctions allow
superconducting currents (intergranular vortices).  Although an
explanation to all details of the temperature dependence of the
remanence shown for example in Figs.~\ref{rem}, \ref{RE-2},
\ref{AU-Rem}, \ref{RE-S2} and \ref{rem-hopg}, requires knowledge
of the pinned entities, the vanishing of $\Delta R(0)$ to lower
temperatures appears at odds with the expected increase in
Josephson coupling, Josephson critical current as well as in the
overall increase of pinning strength  of intragranular vortices
decreasing temperature. Because this behavior measured by the
magnetization\cite{schcar} and resistance appears to be rather
general, we provide below a possible general explanation.

Assuming that the superconductivity has a 2D character and it is
localized at the interfaces, one should first clarify to what
extent this would be at all possible. Following early theoretical
work on this issue (see chap.~7 in Ref.~\cite{sim94}) a
superconducting layer of circular size $R$ (in our case the size
of the 2D interfaces in graphite $R \lesssim 10~\mu$m) can exhibit
superconductivity if $R \ll \lambda_{\perp}$, the effective
screening length for transverse magnetic field. Taking into
account that in 2D the effective London penetration depth is given
by the Pearl result\cite{pea64} $\lambda_{\perp} =
2\lambda_L^2/d$, where $\lambda_L$ is the London penetration depth
of the bulk superconductor ($\lambda_L \sim 1~\mu$m) and $d \sim
1~$nm the thickness of the superconducting layer, we estimate
$\lambda_{\perp} \gtrsim 1$~mm $\gg R$. The existence of Pearl
vortices with such a giant effective penetration depth would imply
a priori a very weak pinning. We speculate therefore that the
trapped flux is mainly due to fluxons, intergranular vortices, and
not intragranular ones. Upon the characteristics of the interfaces
and decreasing temperature, the trapped flux can increase due to
an increase in the Josephson coupling between the superconducting
domains. On the other hand,  decreasing temperature the size of
the superconducting domains may increase, reducing the effective
superconducting loop areas for fluxons. This can be the reason for
the vanishing of trapped flux at $T < 200~$K (see Fig.~\ref{rem}),
in the same temperature range where the giant MIT is observed, see
Fig.~\ref{mit}(a). Our findings would imply then that the MIT has
a relation to superconductivity, as has been speculated in the
past\cite{esqpip}. It is interesting to note that in spite of a
nearly 100 times larger magnetoresistance at 5~K than at $\sim
300$~K (virgin state), see Fig.\ref{ir}(a), it shows basically the
same field dependence in the low field range. The observed time
relaxation of the magnetoresistance after applying a magnetic
field and its change with temperature (see Fig.~\ref{AU-time}) are
also compatible with the existence of vortices/fluxons.

The data presented in this study took more than one and a half
years of systematic measurements in several samples. During the
measurements and after temperature cycling to 390~K some of the
samples several times, we recognized that the strength of the
remanence decreased systematically, although the transition
temperature did not change within experimental error. Note, for
example, that the maximum remanence shown in Fig.~\ref{rem},
measured after several temperature cycles, is $< 7~\mu\Omega$,
whereas the same sample at the beginning of the measurements
showed already at 300~K a remanence of $\sim 40~\mu\Omega$, see
Fig.~\ref{ir}. This relatively low temperature annealing effect
has some similarity to the irreversible behavior observed in the
magnetization of a HOPG sample with interfaces after heating it
above 400~K\cite{schcar}. Because the remanence depends on the
pinning strength and this on the characteristics of the
superconducting regions (size, defects, etc.), it appears
plausible that this low-temperature annealing influences the
remanence and should be taken into account in future experiments.
Note that the domains exhibiting rhombohedral stacking in thin
graphite flakes are stable to $800~^\circ$C~\cite{LUI} and to
$1000~^\circ$C in bulk graphite\cite{MATU,FREIS}. Therefore,
annealing effects appear to be more related to the annealing of
certain defects or even hydrogen diffusion than structural changes
at the interfaces.

The existence of a superconducting transition in graphite  samples
at such high temperatures clarifies to some extent the origin of
the superconducting-like behavior of different graphite samples
reported in the last 40 years\cite{esqpip}. Those reports were not
taken seriously by most of the scientific community. Future
studies should clarify the origin of the superconductivity and the
characteristics of the trapped flux or pinned entities using
methods that allow, for example, its local visualization at the
sample surface.





\acknowledgements P.D.E. gratefully acknowledges fruitful
discussions with: J. G. Rodrigo, also for kindly sharing his
 unpublished low-temperature STM results on graphite;
 T. Heikkil\"a and G. Volovik for explaining their thoughts about
 high-temperature superconductivity at the interfaces
 of graphite and the flat-band effects. A.C. acknowledges the Nacional de Grafite Ltda.,
Itapecerica - MG in Brazil for
 supplying the graphite samples. C.E.P. gratefully acknowledges
 the support provided by The Brazilian National Council for the Improvement of
 Higher Education (CAPES).

\bibliographystyle{apsrev4-1}


%

\end{document}